# Cluster Vertex Deletion Problems on Cubic Graphs

Irena Rusu*

May 12, 2025

### Abstract

The problems CLUSTER VERTEX DELETION (or CLUSTER-VD) and its generalization $s$-CLUB CLUSTER VERTEX DELETION (or $s$-CLUB-VD, for any integer $s \geq 1$), have been introduced with the aim of detecting highly-connected parts in complex systems. Their NP-completeness has been established for several classes of graphs, but remains open for smaller classes, including subcubic planar bipartite graphs and cubic graphs. In this paper, we show that CLUSTER-VD and more generally $s$-CLUB-VD are NP-complete for cubic planar bipartite graphs. We also deduce new results for the related $k$-PATH VERTEX COVER problem (or $k$-PVC), namely 3-PVC is NP-complete for cubic planar bipartite graphs, whereas $k$-PVC with $k \geq 4$ is NP-complete for subcubic planar (and bipartite, when $k$ is odd) graphs of arbitrarily large girth.

## 1 Introduction

Vertex deletion problems ask to remove from a given graph $\mathsf{G} = (V, E)$ a minimum size set $D$ of vertices such that the resulting graph has a given property. The properties we consider here concern the connected components of the remaining graph, that should be highly connected, as required by numerous applications in computational biology [2, 23] and in robustness of real-world networks [1]. When the connected components are asked to be singletons, we obtain the VERTEX COVER problem. When they are asked to contain no (induced or not) $k$-path, we get the $k$-PATH VERTEX COVER problem. Here we deal more particularly with two other problems, called CLUSTER VERTEX DELETION and $s$-CLUB CLUSTER VERTEX DELETION, that have received a lot of attention in the last years. Specific applications for these problems may be found in [15] and [22] respectively.

Given an undirected graph $\mathsf{G} = (V, E)$, a *CVD-set* is a set $D \subseteq V$ such that each connected component of the subgraph induced in $\mathsf{G}$ by $V \setminus D$ is a clique (also called a *cluster* in this context).

---

CLUSTER VERTEX DELETION (CLUSTER-VD)[13]
**Input:**      A graph $\mathsf{G} = (V, E)$, a positive integer $p$.
**Question:**   Is there a CVD-set $D$ of $\mathsf{G}$ of cardinality at most $p$?

---

CLUSTER-VD is the particular case with $s = 1$ of the following more general problem, defined for each integer $s \geq 1$. An *$s$-club* of $\mathsf{G}$ is an induced subgraph of $\mathsf{G}$ with diameter at most $s$. An *$s$-club set* [5] of $\mathsf{G}$ is a set $D \subseteq V$ such that each connected component of the subgraph induced in $\mathsf{G}$ by $V \setminus D$ is an $s$-club.

---

$s$-CLUB CLUSTER VERTEX DELETION ($s$-CLUB-VD)[22]
**Input:**      A graph $\mathsf{G} = (V, E)$, a positive integer $p$.
**Question:**   Is there an $s$-club set $D$ of $\mathsf{G}$ of cardinality at most $p$?

---

The NP-hardness of CLUSTER-VD and $s$-CLUB-VD follows from [20], since both problems concern heredi­tary properties. Focusing on particular classes of graphs, polynomial algorithms exist for CLUSTER-VD on interval

*Nantes Université, École Centrale Nantes, CNRS, LS2N, UMR 6004, F-44000 Nantes, France; email: Irena.Rusu@univ-nantes.fr

[8, 10], split [8], trapezoid [9] as well as well-partitioned chordal graphs [10]. But CLUSTER-VD is also known to be NP-complete on planar bipartite graphs of degree at most 4 [16] and on subcubic planar graphs [14]. Parameterized algorithms for the general case are proposed in [7, 13, 15, 4, 24]. On the (in)approximability side, it is known that CLUSTER-VD is APX-complete in bipartite graphs [16, 17]. Concerning $s$-CLUB-VD ($s \geq 2$), the problem is polynomial when reduced to trapezoid graphs [9] and to interval graphs [10], and NP-complete for planar bipartite graphs with maximum degree 7 [9], as well as for well-partitioned chordal graphs [10]. Parameterized algorithms for the general case have been proposed in [22, 21]. Finally, it is known that $s$-CLUB-VD ($s \geq 2$) is APX-complete on bipartite graphs with maximum degree 7 [9]. These results show that the problems we are interested in have been well studied, but the boundaries between polynomial and hard cases are yet to be determined. In this paper, we improve the above-mentioned results about the hardness of these problems, and prove that CLUSTER-VD and $s$-CLUB-VD ($s \geq 2$) are NP-complete for cubic planar bipartite graphs.

The problem $k$-PATH VERTEX COVER (or $k$-PVC) [6, 26] is related to the abovementioned problems. It asks, given a graph $\mathsf{G}$, to find a minimum vertex set $S$ that has an element in each $k$-path of $\mathsf{G}$, *i.e.* path on $k$ vertices, induced or not. For $k \geq 3$, the problems $(k-2)$-CLUB-VD and $k$-PVC are equivalent for the graphs where each $k$-path is an induced path, *i.e.* for graphs with girth at least $k+1$. Our results allow to deduce that:

- 3-PVC is NP-complete for cubic planar bipartite graphs.
- $k$-PVC with $k \geq 4$ is NP-complete for subcubic planar (and bipartite, when $k$ is odd) graphs of arbitrarily large girth.

**Organization** In Section 2 we give the formal definitions and notations. In Section 3, we present the construction of a graph $\mathsf{G}$, which is the basis of all our reductions, from an instance of (2,3)-PLANAR-3DM. The graph $\mathsf{G}$ is used, after appropriate modifications, in Section 4 and Section 5 to obtain intermediate results on $s$-CLUB-VD ($s \geq 1$, thus including CLUSTER-VD) in subcubic planar bipartite graphs. In Section 6 we transform the subcubic instances into cubic instances, and prove our main theorem on cubic planar bipartite graphs. Section 7 is a short section deducing new results on the $k$-PVC problem, whereas Section 8 is the conclusion.

## 2 Preliminaries

All the graphs we consider are undirected and simple. We denote by $V(\mathsf{G})$ the vertex set and $E(\mathsf{G})$ the edge set of the graph $\mathsf{G}$. Given an integer $k$, we use the notations $P_k$ ($k \geq 1$) and $C_k$ ($k \geq 3$) for the induced paths and cycles with $k$ vertices. A $P_k$ ($C_k$ respectively) with vertices $x_1, x_2, \ldots x_k$ is written as $x_1 x_2 \ldots x_k$ (respectively $x_1 x_2 \ldots x_k x_1$). The *distance* between two vertices is the length, in number of edges, of the shortest path between them. The *girth* of a graph is the size of its shortest induced cycle.

For $x \in V$, we denote the neighborhood of $x$ in $\mathsf{G}$ by $N_{\mathsf{G}}(x)$. An edge between two vertices $x$ and $y$ is written as $xy$. The *subdivision* of an edge $xy$ using $k$ vertices ($k \geq 0$) is the operation that replaces the edge $xy$ with a $P_{k+2}$ whose $k$ vertices of degree 2 are new. The set of 2-degree vertices of the path (also called *subdivision vertices* in the subsequent) is denoted by $SV(x, y)$.

Furthermore, we say that a subset $D$ of $V(\mathsf{G})$ *hits* another subset $V'$ of $V(\mathsf{G})$ whenever $D \cap V' \neq \emptyset$. When $v \in D \cap V'$, we also say that $v$ hits $V'$. The subset $D$ *hits* a subgraph of $\mathsf{G}$ if it hits the vertex set of the subgraph. It is easy to see that a set $D$ is a CVD-set (an $s$-club set respectively) of $\mathsf{G}$ if and only if $D$ hits each $P_3$ ($P_{s+2}$ respectively) of $\mathsf{G}$.

As CLUSTER-VD and 1-CLUB-VD are equivalent, in the remainder of the paper we refer to them both as to $s$-CLUB-VD with $s \geq 1$.

We aim at proving the following results.

**Theorem 1.** *Let $s \geq 1$ and $g > 0$ be two integers. When $s$ is odd, the problem $s$-CLUB-VD is NP-complete for subcubic planar bipartite graphs of girth larger than $g$.*

**Theorem 2.** *Let $s \geq 2$ be an integer. When $s$ is even, the problem $s$-CLUB-VD is NP-complete for subcubic planar bipartite graphs of girth equal to $s + 2$.*

**Theorem 3.** *Let $s \geq 1$ be an integer. The problem $s$-CLUB-VD is NP-complete for cubic planar bipartite graphs.*

These results improve the existing bounds of hardness known for these problems (see Section 1). Note that the NP-completeness of 1-CLUB VD (i.e. CLUSTER-VD) on subcubic planar bipartite graphs has been considered in [14], but we think that the instances of CLUSTER-VD built by the reduction are usually not planar. We discuss this point in Remark 1 at the end of Section 4, and provide here another proof, with a reduction from another NP-complete problem.

The problem that we use is the following one.

---

PLANAR 3-DIMENSIONAL MATCHING (PLANAR-3DM)

**Input:** Three disjoint sets $R$, $B$, and $Y$ such that $|R| = |B| = |Y|$, and a set of triplets $T \subseteq R \times B \times Y$ such that the bipartite graph $\mathsf{H}$ is planar, where

$$V(\mathsf{H}) = T \cup R \cup B \cup Y \text{ and } E(\mathsf{H}) = \bigcup_{\substack{t \in T \\ t=(r,b,y)}} \{tr, tb, ty\}.$$

**Question:** Is there a *3D-matching* $M \subseteq T$, i.e. a subset $M$ of $T$ such that each element of $R \cup B \cup Y$ appears in exactly one element of $M$?

---

It is shown in [12] that PLANAR-3DM is NP-complete even when each $w$ from $R \cup B \cup Y$ occurs in two or three triplets of $T$. We use this restricted variant of the problem, that we denote by (2,3)-PLANAR-3DM. More precisely, our reductions to $s$-CLUB-VD ($s \geq 1$) use the graphs given in the diagram below, built from each other as indicated. The initial graph $\mathsf{H}$ is the bipartite planar graph obtained from an instance of (2,3)-PLANAR-3DM.

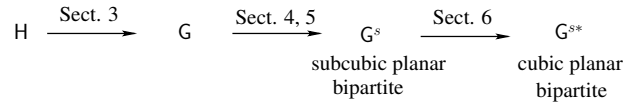

Graphs and sets are denoted by capital letters with distinct fonts, like $\mathsf{T}$ and $T$ respectively. Some exceptions exist however, for the induced paths, induced cycles, and some particular sets. Elements of a set and more particularly vertices are denoted with lower case letters.

## 3 Main construction

Consider an instance of (2,3)-PLANAR-3DM: three disjoint sets $R$, $B$, and $Y$ such that $|R| = |B| = |Y|$ and a set of triplets $T \subseteq R \times B \times Y$. We let $W = R \cup B \cup Y$, and assume that each element $w$ of $W$ occurs in two or three triplets from $T$. As before, we define the bipartite graph $\mathsf{H}$ as follows:

$$V(\mathsf{H}) = T \cup W \text{ and } E(\mathsf{H}) = \bigcup_{\substack{t \in T \\ t=(r,b,y)}} \{tr, tb, ty\}.$$

A vertex $w$ from $W$ is of *type 2* if its degree in $\mathsf{H}$ is 2 and of *type 3* if its degree in $\mathsf{H}$ is 3. We denote by $n_2$ ($n_3$ respectively) the number of vertices of $W$ of type 2 (3 respectively) in $\mathsf{H}$. Then $n_2 + n_3 = |W|$.

By hypothesis, $\mathsf{H}$ is planar. Consider a planar embedding $\mathcal{H}$ of $\mathsf{H}$ in the plane, and define the graph $\mathsf{G}$ as follows.

- replace each vertex $t$ (with $t \in T$) of $\mathsf{H}$ with a $T$**-gadget** $\mathsf{T}$ built as in Figure 1. (Note the font difference between $T$ and $\mathsf{T}$.) The $T$-gadget $\mathsf{T}$ is a cycle. Each neighbor $w$ of $t$ in $\mathsf{H}$ is represented by the following path, called the *support* of $w$ on $\mathsf{T}$, whose vertices appear in counterclockwise order on $\mathsf{T}$:

$$\mathsf{Supp}_{\mathsf{T}}(w) : \overset{\circ}{s_w} \overset{\bullet}{s_w} \overset{}{s_w} \overset{\circ}{u_w} \overset{\bullet}{u_w} \overset{}{u_w} \overset{\circ}{x_w} \overset{\bullet}{x_w} \overset{}{x_w} \overset{\circ}{z_w} \overset{\bullet}{z_w} \overset{}{z_w}$$

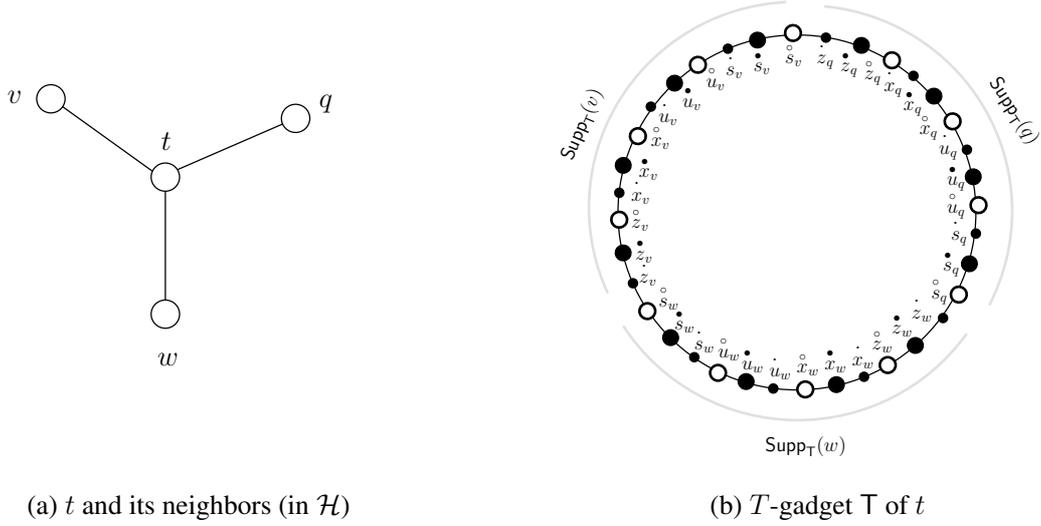

(a) $t$ and its neighbors (in $\mathcal{H}$)

(b) $T$-gadget $\mathsf{T}$ of $t$

Figure 1: (a) Vertex $t$ of $\mathsf{H}$, in the planar embedding $\mathcal{H}$ of $\mathsf{H}$ in the plane. Here, $\{w, q, v\}$ is the set of elements of the triplet $t$. (b) $T$-gadget $\mathsf{T}$ corresponding to $t$. The counterclockwise order of $w, q, v$ in $\mathcal{H}$ is preserved in $\mathsf{T}$, where each neighbor of $t$ is represented by its support.

When $t_i$ from $T$ is used instead of $t$, its $T$-gadget is denoted by $T_i$ and the superscript $i$ is added to each vertex of the support.

- replace each vertex $w$ of $\mathsf{H}$ of type 2, say with $N_{\mathsf{H}}(w) = \{t_i, t_j\}$, with a **2-gadget** as shown in Figure 2a. The 2-gadget is connected with the supports of $w$ on $\mathsf{T}_i$ and $\mathsf{T}_j$ as shown in Figure 2b. Each connected component of a 2-gadget, which is a $P_3$, is called a *chord*. All the 2-gadgets are vertex-disjoint.

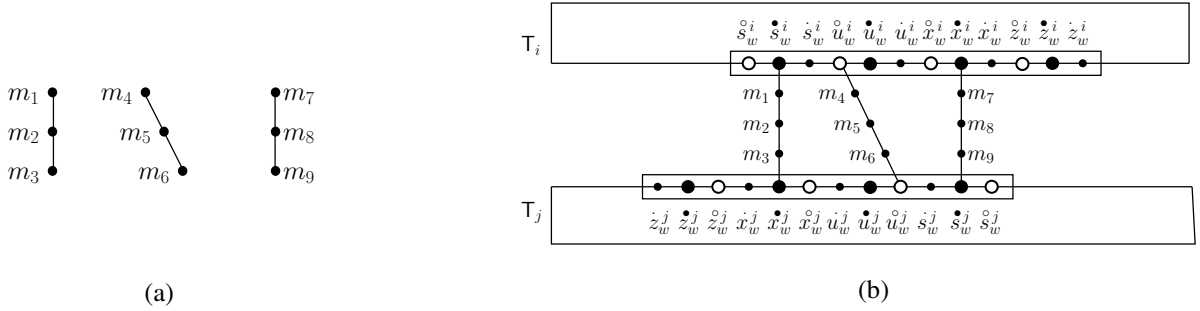

(a)

(b)

Figure 2: (a) The 2-gadget. (b) Its connections with the two supports of $w$ on $\mathsf{T}_i$ and $\mathsf{T}_j$.

- replace each vertex $w$ of $\mathsf{H}$ of type 3, say with $N_{\mathsf{H}}(w) = \{t_i, t_j, t_k\}$, with a **3-gadget** (Figure 3a) connected with $\mathsf{T}_i$, $\mathsf{T}_j$ and $\mathsf{T}_k$ as shown in Figure 3b. Each connected component of a 3-gadget that is a $P_3$ is called a *chord*. The remaining connected component is called the *star* of the 3-gadget. All the 3-gadgets are vertex-disjoint.

Note that the 2- and 3-gadgets are connected with the same vertices of $\mathsf{Supp}_{\mathsf{T}}(w)$, namely with $\overset{\bullet}{s}_w$, $\overset{\circ}{u}_w$, $\overset{\bullet}{x}_w$. Each $v \in \{\overset{\bullet}{s}_w, \overset{\circ}{u}_w, \overset{\bullet}{x}_w\}$ is called a *junction vertex*, and its neighbor in the 2- or 3-gadget connected with it is its *junction neighbor*, denoted $J(v)$.

We define three sets $\mathcal{C}, \mathcal{D}, \mathcal{E}$, called *colors*, containing all the vertices represented and denoted by a circle, a disc and a point respectively. Thus:

$$\mathcal{C} = \bigcup_{\substack{t \in T, t=(r,b,y) \\ w \in \{r,b,y\}}} \{\overset{\bullet}{s}_w, \overset{\circ}{u}_w, \overset{\circ}{x}_w, \overset{\circ}{z}_w\}, \quad \mathcal{D} = \bigcup_{\substack{t \in T, t=(r,b,y) \\ w \in \{r,b,y\}}} \{\overset{\bullet}{s}_w, \overset{\bullet}{u}_w, \overset{\bullet}{x}_w, \overset{\bullet}{z}_w\}, \quad \mathcal{E} = \bigcup_{\substack{t \in T, t=(r,b,y) \\ w \in \{r,b,y\}}} \{\dot{s}_w, \dot{u}_w, \dot{x}_w, \dot{z}_w\}.$$

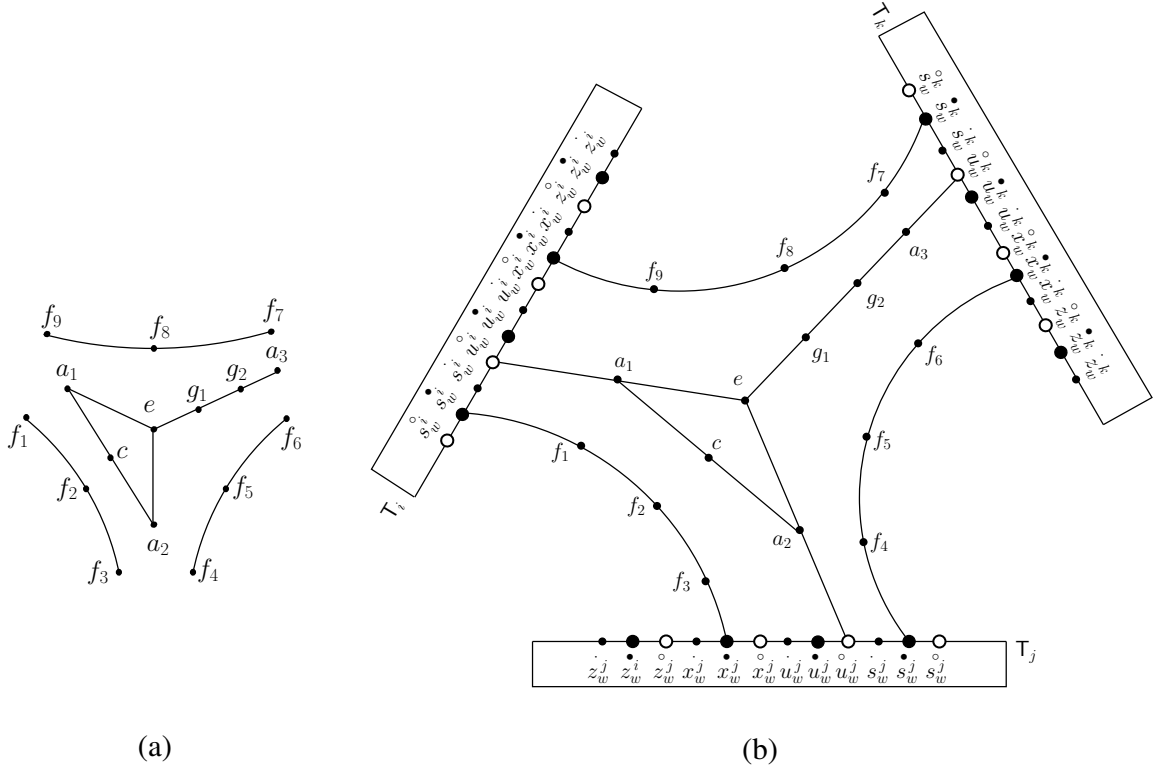

(a)                        (b)

Figure 3: (a) The 3-gadget. (b) Its connections with the three supports of $w$ on $\mathsf{T}_i$, $\mathsf{T}_j$ and $\mathsf{T}_k$.

For a $T$-gadget $\mathsf{T}$ and a color $\mathcal{F} \in \{\mathcal{C}, \mathcal{D}, \mathcal{E}\}$, we let $\mathcal{F}(\mathsf{T}) = V(\mathsf{T}) \cap \mathcal{F}$. We say that $\mathsf{T}$ is $\mathcal{F}$-*colored* by a set $D$ of vertices from $\mathsf{G}$ if $D \cap V(\mathsf{T}) = \mathcal{F}(\mathsf{T})$.

In our proofs, we appropriately modify $\mathsf{G}$ - that is built from the instance $\mathsf{H}$ of $(2,3)$-PLANAR-3DM - to obtain an instance of $s$-CLUB-VD, for each $s \geq 1$.

# 4   Proof of Theorem 1

The theorem we prove in this section is:

**Theorem 1.** *Let $s \geq 1$ and $g > 0$ be two integers. When $s$ is odd, the problem $s$-CLUB-VD is NP-complete for subcubic planar bipartite graphs of girth larger than $g$.*

Let $s \geq 1$ and odd. We first reduce $(2,3)$-PLANAR-3DM to $s$-CLUB-VD for subcubic planar bipartite instances of girth at least $s + 3$, then show how to appropriately transform these instances into instances with girth larger than $g$. To this end, we modify the graph $\mathsf{G}$ built in Section 3 by performing subdivisions that always use new vertices, by definition.

## 4.1   Graphs with girth at least $s + 3$

Let $\mathsf{G}^s$ be the graph obtained from $\mathsf{G}$ by performing the following operations:

- for each $T$-gadget $\mathsf{T}$ (see Figure 1b), subdivide using $s - 1$ vertices each of the edges $xy$ such that $x \in \mathcal{E}(\mathsf{T})$ and $y \in \mathcal{C}(\mathsf{T})$.

- for each 2-gadget (see Figure 2a), subdivide using $s - 1$ vertices each of the edges $m_1 m_2, m_4 m_5, m_7 m_8$.

- for each 3-gadget (see Figure 3a), subdivide using $s - 1$ vertices each of the edges $f_1 f_2, f_4 f_5, f_7 f_8, g_1 g_2$, then subdivide each of the edges $a_1 e, a_2 e$ using $(s-1)/2$ vertices when $(s-1)/2$ is even and using $(s-1)/2 + (s+2)$ vertices otherwise.

The vertices (edges) of $\mathsf{G}^s$ that were already present in $\mathsf{G}$ are called *standard vertices (edges)*. The vertices added by the subdivisions are called *subdivision vertices*. Recall that $SV(x, y)$ denotes the set of subdivision vertices resulting from the subdivision of the standard edge $xy$ (see Section 2). The support of $w$ (for $w \in W$) in some $T$-gadget $\mathsf{T}$ of $\mathsf{G}^s$ contains, in addition to the standard vertices, the sets $SV(\dot{s}_w, \overset{\circ}{u}_w)$, $SV(\dot{u}_w, \overset{\circ}{x}_w)$, $SV(\dot{x}_w, \overset{\circ}{z}_w)$ and $SV(\dot{z}_w, \overset{\circ}{s}_{w'})$, where $w'$ is the element of $W$ whose support comes next along $T$ in the counterclockwise direction. Consequently, in $\mathsf{G}^s$, $\mathsf{Supp}_\mathsf{T}(w)$ has $12 + 4(s-1) = 4s + 8$ vertices.

The sets $\mathcal{C}, \mathcal{D}, \mathcal{E}$ are unchanged, as well as $\mathcal{C}(\mathsf{T}), \mathcal{D}(\mathsf{T}), \mathcal{E}(\mathsf{T})$ for any $T$-gadget $\mathsf{T}$. Moreover, we define $\mathcal{E}_q(\mathsf{T})$ ($1 \leq q \leq s-1$) to be the set that contains the $q$-th subdivision vertex of each $SV(x, y)$ from $\mathsf{T}$, when going from $x$ to $y$ in counterclockwise direction. Then $\mathcal{E}_q = \bigcup_{t \in T} \mathcal{E}_q(\mathsf{T})$. We also rename $\mathcal{E}$ as $\mathcal{E}_0$, and remark that $\mathsf{T}$ is made of twelve disjoint $P_{s+2}$ paths whose vertices, when considered in counterclockwise direction, have the colors $\mathcal{C}, \mathcal{D}, \mathcal{E}_0, \ldots, \mathcal{E}_{s-1}$ respectively, in this order.

The following claims state properties of our construction.

**Claim 4.1.** *Let $s \geq 1$ be an odd integer. The graph $\mathsf{G}^s$ is a subcubic planar bipartite graph of girth at least $s + 3$.*

*Proof.* The planarity and maximum degree of $\mathsf{G}^s$ are immediate. When at least one 3-gadget exists in $\mathsf{G}^s$, the girth is given by the cycle of its star, induced by $\{a_1, SV(a_1, e), e, SV(a_2, e), a_2, c\}$, which has $2 \cdot (s-1)/2 + 4 = s + 3$ vertices when $(s-1)/2$ is even, and $s + 3 + 2(s+2) = 3s + 7$ vertices otherwise. If no 3-gadget exists, the girth is given by each induced ycle containing two chords of a 2-gadget and the two paths joining them in $\mathsf{T}_i$ and $\mathsf{T}_j$. The girth is then $4(s-1) + 14 = 4s + 10$.

We prove that $\mathsf{G}^s$ is bipartite. All the gadgets are bipartite, so we consider: the bipartition $(X_1, Y_1)$ of each $T$-gadget $\mathsf{T}$ such that $\overset{\circ}{s}_w \in X_1$, the bipartition $(X_2, Y_2)$ of each 2-gadget such that $m_1, m_4, m_7 \in X_2$ and the bipartition $(X_3, Y_3)$ of each 3-gadget such that $f_1, f_4, f_7, a_1 \in X_3$.

The graph $\mathsf{G}^s$ is bipartite with the following bipartition:

$$V_1 = \bigcup_{\substack{\text{all} \\ T-\text{gadgets in } \mathsf{G}^s}} X_1 \quad \cup \quad \bigcup_{\substack{\text{all} \\ 2\text{-gadgets in } \mathsf{G}^s}} X_2 \quad \cup \quad \bigcup_{\substack{\text{all} \\ 3\text{-gadgets in } \mathsf{G}^s}} X_3$$

$$V_2 = \bigcup_{\substack{\text{all} \\ T-\text{gadgets in } \mathsf{G}^s}} Y_1 \quad \cup \quad \bigcup_{\substack{\text{all} \\ 2\text{-gadgets in } \mathsf{G}^s}} Y_2 \quad \cup \quad \bigcup_{\substack{\text{all} \\ 3\text{-gadgets in } \mathsf{G}^s}} Y_3$$

$\square$

**Claim 4.2.** *Let $s \geq 1$ be an odd integer. The following affirmations hold:*

(a) *Each $T$-gadget $\mathsf{T}$ admits exactly $s + 2$ minimum $s$-club sets: $\mathcal{C}(\mathsf{T}), \mathcal{D}(\mathsf{T}), \mathcal{E}_0(\mathsf{T}), \mathcal{E}_1(\mathsf{T}), \ldots, \mathcal{E}_{s-1}(\mathsf{T})$. Their size is 12.*

(b) *The minimum $s$-club sets of a 2-gadget are of size 3.*

(c) *The minimum $s$-club sets of a 3-gadget are of size 6 when $(s-1)/2$ is even, and of size 8 otherwise.*

*Proof.* Each $T$-gadget $\mathsf{T}$ is made of 12 disjoint subpaths of $s + 2$ vertices each, whose colors follow the same order when the vertices are considered in counterclockwise direction. Each subpath needs to be hit by any $s$-club set $D$ using at least one vertex, and two vertices of $D$ that are consecutive along $\mathsf{T}$ must be at distance at most $s + 2$ (otherwise a $P_{s+2}$ non-hit by $D$ exists between them). Then affirmation (a) follows.

In each 2-gadget, each chord is a $P_{s+2}$ and the conclusion follows.

In each 3-gadget, the same affirmation holds for the three chords, as well as for the path joining $g_1$ to $a_3$. The cycle of the star is of size $s + 3$ when $(s-1)/2$ is even, and of size $3s + 7$ otherwise. We deduce that this cycle must be hit by at least 2, respectively at least 4, vertices of any $s$-club set $D$, which implies that overall $D$ must have at least 6, respectively 8, vertices. Moreover, it may be easiclu checked that $\{f_1, f_4, f_7, a_1, a_3, e\}$ is an $s$-club set of size 6 in the former case. In the latter case, each path between $a_1, e$ and between $a_2, e$ requires to add another vertex to the $s$-club set, due to the extra $s + 2$ vertices (that were added in order to obtain a bipartite graph). $\square$

Let $F(s) = 6$ when $(s-1)/2$ is even, and $F(s) = 8$ otherwise. Recalling that $n_2, n_3$ denote the number of vertices of type 2 and 3 respectively in $\mathsf{H}$, we deduce from Claim 4.2:

**Corollary 4.3.** *Let $s \geq 1$ be an odd integer. Each $s$-club set of $\mathsf{G}^s$ has size at least $12|T| + 3n_2 + F(s)n_3$.*

As the problem $s$-Club-VD is obviously in NP, we need to show that the construction that starts with a bipartite graph $\mathsf{H}$ representing an instance of (2,3)-Planar-3DM, builds $\mathsf{G}$ and then builds $\mathsf{G}^s$ (with odd $s$) is a reduction from (2,3)-Planar-3DM to $s$-Club-VD for subcubic planar bipartite graphs.

**Claim 4.4.** *Let $s \geq 1$ be an odd integer. There exists a 3D-matching $M \subseteq T$ if and only if $\mathsf{G}^s$ contains an $s$-club set of size equal to $12|T| + 3n_2 + F(s)n_3$.*

*Proof.* The intuitive idea behind the proof is that a 3D-matching exists if and only if there is an $s$-club set $D$ of $\mathsf{G}^s$ that $\mathcal{C}$-colors all the $T$-gadgets corresponding to elements of the 3D-matching, and $\mathcal{D}$-colors all the other $T$-gadgets.

$\Rightarrow$: Let $M$ be a 3D-matching. Consider, in $\mathsf{G}^s$, the set $D$ built as follows ($D$ is initially empty):

- for each $t \in M$, add the elements of $\mathcal{C}(\mathsf{T})$ to $D$.

- for each $t \in T \setminus M$, add the elements of $\mathcal{D}(\mathsf{T})$ to $D$.

- for each vertex $w \in W$ of type 2 adjacent to $t_i$ and $t_j$ in $\mathsf{H}$ (see Figure 2b):

    ⋆ if $t_i \in M$ then add to $D$ the set $\{m_1, m_6, m_7\}$

    ⋆ if $t_j \in M$ then add to $D$ the set $\{m_3, m_4, m_9\}$.

- for each vertex $w \in W$ of type 3 adjacent to $t_i, t_j$ and $t_k$ in $\mathsf{H}$ (see Figure 3b):

    ⋆ if $t_i \in M$ then add to $D$ the set $\{f_1, f_5, f_9, a_2, a_3, e\}$. Moreover, when $(s-1)/2$ is odd, add to $D$ the vertex of $SV(a_1, e)$ that is at distance $s$ from $a_1$ and the vertex of $SV(a_2, e)$ that is at distance $s$ from $a_2$.

    ⋆ if $t_j \in M$ then add to $D$ the set $\{f_3, f_4, f_8, a_1, a_3, e\}$. Moreover, when $(s-1)/2$ is odd, add to $D$ the vertex of $SV(a_1, e)$ that is at distance $s$ from $a_1$ and the vertex of $SV(a_2, e)$ that is at distance $s$ from $a_2$.

    ⋆ if $t_k \in M$ then add to $D$ the set $\{f_2, f_6, f_7, a_1, a_2, g_1\}$. Moreover, when $(s-1)/2$ is odd, add to $D$ the vertex of $SV(a_1, e)$ that is at distance $s+2$ from $a_1$ and the vertex of $SV(a_2, e)$ that is at distance $s+2$ from $a_2$.

By Claim 4.2a, each $P_{s+2}$ belonging to some $T$-gadget is hit by $D$. The definition of $D$ easily implies that each $P_{s+2}$ contained in a 2- or 3-gadget is also hit by $D$. Moreover, for each junction vertex $v$, either $v$ or $J(v)$ belongs to $D$ by definition, implying that each $P_{s+2}$ with vertices from both a $T$-gadget and a 2- or 3-gadget is also hit by $D$.

By construction, the cardinality of $D$ is $12|T| + 3n_2 + F(s)n_3$.

$\Leftarrow$: Let $D$ be an $s$-club set of $\mathsf{G}$, with cardinality $12|T| + 3n_2 + 6n_3$. Then $D$ has the following four properties, that we subsequently use to deduce the colors of the $T$-gadgets connected with a 2- or 3-gadget:

(a) $D$ hits:

    ($a_1$) each $T$-gadget by exactly 12 vertices, those of a unique color among $\mathcal{C}, \mathcal{D}$ and $\mathcal{E}_q$ ($0 \leq q \leq s-1$).

    ($a_2$) each chord of a 2- or 3-gadgets by exactly one vertex.

    ($a_3$) the star of each 3-gadget by exactly 3 vertices when $(s-1)/2$ is even, and with exactly 5 vertices otherwise.

*Proof.* By Corollary 4.3, $D$ has the minimum possible size. By Claim 4.2, this size can only be obtained when the restriction of $D$ to each gadget is an $s$-club set for the gadget. Then affirmation ($a_1$) follows from Claim 4.2a, affirmation ($a_2$) is immediately deduced since each chord is a $P_{s+2}$ thus needs to be hit by at least one vertex and Claim 4.2 allows no more than one vertex for each chord, and affirmation ($a_3$) follows from ($a_2$) and Claim 4.2c.

(b) In a 3-gadget, we cannot have simultaneously $a_1, a_2, a_3 \in D$.

*Proof.* By contradiction, we assume (see Figure 3a) that $a_1, a_2, a_3 \in D$. Since, by affirmation $(a_3)$, $D$ can contain only three vertices of the star (five respectively when $(s-1)/2$ is odd, two of which belong to $SV(a_1, e)$ and $SV(a_2, e)$), we deduce that the path induced by $SV(g_1, g_2) \cup \{e, g_1, g_2\}$ is a $P_{s+2}$ that is not hit, a contradiction.

(c) Each junction vertex $v$ not belonging to $D$ is the endpoint, in the $T$-gadget $\mathsf{T}$ containing it, of a path non-hit by $D$, which is:

($c_1$) a $P_{s+1}$ if $\mathsf{T}$ is either $\mathcal{C}$- or $\mathcal{D}$-colored by $D$.

($c_2$) a $P_h$ with $h \geq (s+3)/2$ when $\mathsf{T}$ is $\mathcal{E}_q$-colored by $D$ $(0 \leq q \leq s-1)$.

The vertex set of this path is denoted by $L(v)$.

*Proof.* To show this, assume that $\mathsf{T}$ is $\mathcal{F}$-colored, with $\mathcal{F} \in \{\mathcal{C}, \mathcal{D}, \mathcal{E}_0, \ldots, \mathcal{E}_{s-1}\}$, and note that between two consecutive vertices colored $\mathcal{F}$ in $\mathsf{T}$ there are $s+1$ intermediate vertices not belonging to $D$. Recall that $v \notin D$. When $\mathcal{F} \in \{\mathcal{C}, \mathcal{D}\}$, one neighbor of $v$ is colored $\mathcal{F}$ and thus $v$ is the endpoint of the $P_{s+1}$ path made of the $s+1$ intermediate vertices. When $\mathcal{F}$ is some $\mathcal{E}_q$, the vertex $v$ is one of the $s+1$ intermediate vertices not belonging to $D$, and is thus the endpoint of two paths of intermediate vertices, one in each direction (clockwise, counterclockwise). The longest of these two paths has minimum length when $v$ is central among the $s+1$ intermediate vertices, that is for instance when $q = (s-1)/2$. Then the number of vertices of the longest path with endpoint $v$, among the left and the right path, is $(s+3)/2$.

(d) For each junction vertex $v$ from a $T$-gadget $\mathsf{T}$, the set $D$ contains:

($d_1$) at least one vertex from the set $\{v, J(v)\}$, when $\mathsf{T}$ that is either $\mathcal{C}$- or $\mathcal{D}$-colored by $D$.

($d_2$) at least one vertex from each $P_{(s+1)/2}$ with endpoint $J(v)$ and the other vertices in the 2- or 3-gadget containing $J(v)$, when $\mathsf{T}$ is $\mathcal{E}_q$-colored by $D$ $(0 \leq q \leq s-1)$.

*Proof.* Affirmation $(d_1)$ immediately follows from affirmation $(c_1)$ since the path $L(v) \cup \{J(v)\}$ has $s+2$ vertices and must be hit by $D$. Affirmation $(d_2)$ follows from the remark that the path induced by $L(v)$, according to affirmation $(c_2)$, and the $(s+1)/2$ vertices of the path starting with $J(v)$ has at least $s+2$ vertices, and thus it must be hit by $D$.

Using these properties, we now show, following the steps below, that each 2- or 3-gadget is connected with exactly one $T$-gadget $\mathcal{C}$-colored by $D$, whereas the other $T$-gadgets connected with the 2- or 3-gadget are $\mathcal{D}$-colored by $D$:

(e) When two $T$-gadgets are connected by the same 2- or 3-gadget, at least one of them is $\mathcal{D}$-colored by $D$.

*Proof.* The proof is by contradiction. Without loss of generality and in order to fix ideas, we assume the two gadgets colored by $\mathcal{C}$ or $\mathcal{E}_q$, $0 \leq q \leq s-1$, are denoted by $\mathsf{T}_i$ and $\mathsf{T}_j$, and we consider the chord (say with standard vertices $m_1, m_2, m_3$) connecting $\overset{\bullet}{s}_w^i$ to $\overset{\bullet}{x}_w^j$ (See Figure 2b; the reasoning is similar on Figure 3b). By affirmations $(c_1)$ and $(c_2)$ we deduce that the path induced by $L(\overset{\bullet}{s}_w^i) \cup \{m_1, m_2, m_3\} \cup SV(m_1, m_2) \cup L(\overset{\bullet}{x}_w^j)$ has at least $2 \cdot (s+3)/2 + 3 + (s-1)$ vertices, that is, at least $2s+5$ vertices. By affirmation $(a_2)$, the chord contained in this path is hit by a unique vertex of $D$, implying that at least one subpath with $s+2$ vertices will not be hit by $D$, a contradiction.

(f) No $T$-gadget is $\mathcal{E}_q$-colored by $D$ $(0 \leq q \leq s-1)$.

*Proof.* We assume by contradiction that a $T$-gadget $\mathsf{T}_i$ is $\mathcal{E}_q$-colored by $D$, and study the different cases.

- If $\mathsf{T}_i$ is connected with a 2-gadget, then by affirmation (e) $\mathsf{T}_j$ is $\mathcal{D}$-colored by $D$ and thus by affirmation $(d_1)$ we deduce that $m_6 \in D$. Then affirmation $(d_2)$ with $v = \overset{\circ}{u}_w^i$ is contradicted since the path induced by $\{m_4, m_5\} \cup SV(m_4, m_5)$ is not hit by $D$. The reasoning is similar when $\mathsf{T}_j$ (instead of $\mathsf{T}_i$) is $\mathcal{E}_q$-colored by $D$.

- If $\mathsf{T}_i$ is connected with a 3-gadget, by affirmation (e) we similarly deduce that $\mathsf{T}_j$ and $\mathsf{T}_k$ are $\mathcal{D}$-colored by $D$. Thus by affirmation $(d_1)$ we have $a_2, a_3 \in D$ and by affirmation (b) we deduce $a_1 \notin D$. Let the set $SV'$ contain the $(s-1)/2$ vertices of $SV(a_1,e)$ at distance at most $(s-1)/2$ from $a_1$. By affirmation $(d_2)$ for $v = \overset{\circ}{u}{}^i_w$, $SV' \cap D \neq \emptyset$. This implies $s > 1$ (since $SV' = \emptyset$ for $s = 1$). Now, $D$ already contains three vertices from the star when $(s-1)/2$ is even (five vertices otherwise), namely $a_2, a_3$ and a vertex of $SV'$ (and two other vertices from $SV(a_1,e) \setminus SV'$ and $SV(a_2,e)$, when $(s-1)/2$ is odd). By affirmation $(a_3)$, $D$ cannot contain more vertices of the star. But then the path joining $g_2$ to $e$ through $SV(g_1,g_2) \cup \{g_1\}$ has $s+2$ vertices and is not hit, a contradiction.

- The proof is similar when $\mathsf{T}_j$, instead of $\mathsf{T}_i$, is $\mathcal{E}_q$-colored by $D$ (by symmetry with $\mathsf{T}_i$), but also when $\mathsf{T}_k$, instead of $\mathsf{T}_i$, is $\mathcal{E}_q$-colored by $D$. In the latter case, we deduce by affirmation (e) that $\mathsf{T}_i$, $\mathsf{T}_j$ are $\mathcal{D}$-colored, thus by affirmations $(d_1)$ and (b) that $a_1, a_2 \in D$ and $a_3 \notin D$. Moreover, let $SV'$ be the empty set when $(s-1)/2$ is even, and let $SV'$ by the set containing the first $s+2$ subdivision vertices from $SV(a_1,e)$ when going from $a_1$ to $e$, when $(s-1)/2$ is odd. Let $SV'' = SV(a_1,e) \setminus SV'$. We then note two facts. Firstly, $SV'' \cap D = \emptyset$. Indeed, the cycle of the star is already hit by the two vertices $a_1, a_2$ and moreover, when $(s-1)/2$ is odd, by one vertex from $SV'$ and another one from $SV(a_2,e)$. This cycle cannot be hit by more than two, respectively four (when $(s-1)/2$ is odd), vertices by affirmation $(a_3)$ and since another vertex of $D$ must exist on the path from $a_3$ to $e$. Secondly, $|SV''| = (s-1)/2$, by the definition of $SV(a_1,e)$. Thus the path induced by $L(\overset{\circ}{u}{}^k_w) \cup \{a_3, g_2, g_1, e\} \cup SV(g_1,g_2) \cup SV''$ has at least $(s+3)/2 + 4 + (s-1) + (s-1)/2 = 2s+4$ vertices and is hit by only one vertex of $D$ (by affirmation $(a_3)$ and the vertices already in $D$), a contradiction.

(g) Each 2- or 3-gadget is connected with at most one $T$-gadget that is $\mathcal{C}$-colored by $D$.
    *Proof.* In the contrary case, by affirmation $(d_1)$ the chord connecting $\overset{\bullet}{s}{}^i_w$ and $\overset{\bullet}{x}{}^j_w$, where $\mathsf{T}_i, \mathsf{T}_j$ are the two $T$-gadgets $\mathcal{C}$-colored by $D$, contains two vertices from $D$, namely $J(\overset{\bullet}{s}{}^i_w)$ and $J(\overset{\bullet}{s}{}^j_w)$. We have a contradiction with $(a_2)$.

(h) No 2-gadget is connected with two $T$-gadgets $\mathcal{D}$-colored by $D$.
    *Proof.* In the contrary case, $\overset{\circ}{u}{}^i_w, \overset{\circ}{u}{}^j_w \notin D$. By affirmation $(d_1)$ we deduce that the chord connecting them is hit by two vertices of $D$, namely $J(\overset{\circ}{u}{}^i_w), J(\overset{\circ}{u}{}^j_w)$, which contradicts $(a_2)$.

(i) No 3-gadget is connected with three $T$-gadgets $\mathcal{D}$-colored by $D$.
    *Proof.* In the contrary case, since $\overset{\circ}{u}{}^i_w, \overset{\circ}{u}{}^j_w, \overset{\circ}{u}{}^k_w \notin D$, by affirmation $(d_1)$ we deduce that $a_1, a_2, a_3 \in D$. Then we have a contradiction with affirmation (b).

Using affirmations (f), (g), (h) and (i) we deduce that each 2- and 3-gadget is connected with exactly one $T$-gadget that is $\mathcal{C}$-colored by $D$, and we define $M$ to be the set containing all the $T$-gadgets that are $\mathcal{C}$-colored by $D$. □

Claim 4.1, Claim 4.4 and Corollary 4.3, together with the observation that $s$-CLUB-VD belongs to NP, imply that $s$-CLUB-VD is NP-complete when reduced to subcubic planar bipartite graphs of girth at least $s+3$.

In order to prove Theorem 1, it remains to show how to reduce $s$-CLUB-VD on subcubic planar bipartite graphs of girth at least $s+3$ to $s$-CLUB-VD on subcubic planar bipartite graphs of arbitrarily large girth. This is done following the ideas in [19], that address this problem in the case $s = 1$. When $s = 1$, subdividing each edge of a graph $\mathsf{F}$ with no triangles using three new vertices results in a graph $\mathsf{F}^+$ which admits a minimum 1-club set (or CVD-set) of size $h + m(\mathsf{F})$ if and only if $\mathsf{F}$ admits a minimum 1-club set of size $h$. The graph $\mathsf{F}^+$ is subcubic, planar and bipartite whenever $\mathsf{F}$ has these properties, and has girth $g^+ = q * 4$, where $q$ is the girth of $\mathsf{F}$. As we need to exceed a given girth $g$ for our instances, it is sufficient to repeat this operation $log_4 g$ times.

When $s > 1$, we need to show that a similar approach is possible. We do this in Subsection 4.2.

Before going further, we make a remark concerning the even values of $s$. The subdivisions performed on $\mathsf{G}$ in order to obtain $\mathsf{G}^s$ when $s$ is odd are not sufficient when $s$ is even. To see this, consider an even integer $s \geq 2$ and build a new graph from $\mathsf{G}$ by performing subdivisions similar to that for odd $s$. More precisely, use subdivisions with $s-1$ vertices in the $T$-gadgets, 2-gadgets, the chords and the edge $g_1 g_2$ of the 3-gadgets, and $s/2$ subdivisons for $a_1 e, a_2 e$ in the 3-gadgets. Then the resulting graph is subcubic and planar but not bipartite, since the chords are odd length paths, and the cycle induced by the three chords of the same 3-gadget together with the paths joining their endpoints through the $T$-gadgets is odd. However, the reasoning we performed in Claim 4.2 (with $s$-club set of size 6 only, in (c)) and Claim 4.4 remain valid since they depend only of the length of the paths we build, not of their parity. We then deduce the following corollary, where the girth is given by the cycle of the 3-gadget, now of size $s + 4$:

**Corollary 4.5.** *When $s \geq 2$ is an even integer, $s$-Club-VD is NP-complete for subcubic planar graphs of girth at least $s + 4$.*

The construction of subcubic planar and moreover bipartite graphs for an even $s$ needs deeper modifications for the gadgets. We postpone it until Section 5.

## 4.2 Graphs with arbitrarily large girth

The following claim generalizes the result proved in [19] for $s = 1$.

**Claim 4.6.** *Consider a graph $\mathsf{F}$ and an arbitrary integer $s \geq 1$. Let $xy$ be an edge of $\mathsf{F}$ which belongs to no cycle $C_r$ with $r \leq s + 2$, and define $\mathsf{Q}$ to be the graph obtained from $\mathsf{F}$ by subdividing $xy$ using $s + 2$ new vertices $e_1, e_2, \ldots, e_{s+2}$. Then a set $D \subseteq V(\mathsf{F})$ is a minimum $s$-club set of $\mathsf{F}$ if and only if there exists $e \in \{e_1, \ldots, e_{s+2}\}$ such that $D \cup \{e\}$ is a minimum $s$-club set of $\mathsf{Q}$.*

*Proof.* We assume that the vertices $e_1, e_2, \ldots, e_{s+2}$ appear in this order on the path from $x$ to $y$.

Let $\mathcal{P}_{s+2}(\mathsf{F})$ be the set of $P_{s+2}$ of $\mathsf{F}$. The hypothesis that $xy$ does not belong to a $C_r$ cycle with $r \leq s + 2$ implies that no $P_{s+2}$ of $\mathsf{Q}$ contains both $x$ and $y$. More precisely:

$$
\begin{aligned}
\mathcal{P}_{s+2}(\mathsf{Q}) = \mathcal{P}_{s+2}(\mathsf{F}) &\setminus \bigcup_{i=0}^{s} \{a_1 \ldots a_i x y a_{i+3} \ldots a_{s+2} \,|\, a_1 \ldots a_i x y a_{i+3} \ldots a_{s+2} \in \mathcal{P}_{s+2}(\mathsf{F})\} \\
&\cup \bigcup_{i=0}^{s} \{a_1 \ldots a_i x e_1 \ldots e_{s-i+1} \,|\, a_1 \ldots a_i x y a_{i+3} \ldots a_{s+2} \in \mathcal{P}_{s+2}(\mathsf{F})\} \\
&\cup \bigcup_{i=0}^{s} \{e_{s-i+2} \ldots e_{s+2} y a_{i+3} \ldots a_{s+2} \,|\, a_1 \ldots a_i x y a_{i+3} \ldots a_{s+2} \in \mathcal{P}_{s+2}(\mathsf{F})\} \\
&\cup \{e_1 \ldots e_{s+2}\}
\end{aligned}
$$

Let $D \subseteq V(\mathsf{F})$. We first show that $D$ is an $s$-club set of $\mathsf{F}$ if and only if there exists $e \in \{e_1, \ldots, e_{s+2}\}$ such that $D \cup \{e\}$ is an $s$-club set of $\mathsf{Q}$. Then we consider the minimality of the $s$-club set.

$\Rightarrow$: Assume that $D$ is an $s$-club set of $\mathsf{F}$. In $\mathsf{Q}$, let $D'$ be the set obtained by adding to $D$ either $e_1$ (when $x \notin D$ and $y \in D$), or $e_{s+2}$ (when $y \notin D$ and $x \in D$), or $e_{\lceil (s+2)/2 \rceil}$ (when $x, y \in D$). Note that in all these cases the resulting set $D'$ is an $s$-club set of $\mathsf{Q}$.

Consider now the case where $x, y \notin D$. For each $P_{s+2}$ $a_1 \ldots a_i x y a_{i+3} \ldots a_{s+2}$ of $\mathsf{F}$ ($0 \leq i \leq s$), put the path $a_1 \ldots a_i$ (if it is not empty) into a set *Left* and the path $a_{i+3} \ldots a_{s+2}$ (if it is not empty) into a set *Right*. We now have three cases:

- if each path from *Left* is hit by $D$, then let $D' = D \cup \{e_{s+2}\}$, and note that $D'$ is an $s$-club set of $\mathsf{Q}$.

- if each path from *Right* is hit by $D$, then let $D' = D \cup \{e_1\}$, and note that $D'$ is an $s$-club set of $\mathsf{Q}$.

- if none of the above is true, let $\mathsf{P}_L$ be the longest path in *Left* which is not hit by $D$, and $\mathsf{P}_R$ be the longest path in *Right* which is not hit by $D$. Then $\mathsf{P}_L$ and $\mathsf{P}_R$ are vertex disjoint and there is no edge between vertices from $V(\mathsf{P}_L)$ and from $V(\mathsf{P}_R)$. Otherwise $V(\mathsf{P}_L) \cup V(\mathsf{P}_R) \cup \{x, y\}$ induce in $\mathsf{F}$ a cycle that must have at least $s + 3$ vertices (by hypothesis) and thus contains a $P_{s+2}$ that is not hit by $D$, a contradiction. Moreover, the path $\mathsf{P}' = \mathsf{P}_L x y \mathsf{P}_R$ of $\mathsf{F}$ is an induced path with less than $s + 2$ vertices, as it is not hit by $D$. Thus the path obtained in $\mathsf{Q}$ from $\mathsf{P}'$ by subdividing $xy$ with the $s + 2$ vertices $e_1, \ldots, e_{s+2}$ is a $P_h$ with $h \leq 2s + 3$. Then the vertex $e_{s+2-|V(\mathsf{P}_L)|-1}$ belongs to all the $P_{s+2}$ subpaths of $\mathsf{P}'$. Given that $\mathsf{P}_L$ and $\mathsf{P}_R$ were chosen to be as long as possible and are not empty, we deduce that $e_{s+2-|V(\mathsf{P}_L)|-1}$ also belongs to all the $P_{s+2}$ subpaths of any path built using a path from *Left*, the vertices $x, e_2, \ldots, e_{s+2}, y$ and a path from *Right*. Thus $D' = D \cup \{e_{s+2-|V(\mathsf{P}_L)|-1}\}$ is an $s$-club set of $\mathsf{Q}$.

$\Leftarrow$: Let $D' = D \cup \{e\}$ be an $s$-club set of $\mathsf{Q}$. Assume $e = e_i$ for some $i$ with $1 \leq i \leq s + 2$. Then $e_j \notin D'$, for each $j \neq i$. Note first that the paths in $\mathcal{P}_{s+2}(\mathsf{F}) \cap \mathcal{P}_{s+2}(\mathsf{Q})$ are hit by $D$ since $D'$ is an $s$-club of $\mathsf{Q}$ and $e$ does not belong to these paths. The paths from $\mathcal{P}_{s+2}(\mathsf{F}) \setminus \mathcal{P}_{s+2}(\mathsf{Q})$ are of the form $\mathsf{P}_L xy \mathsf{P}_R$, where each of the subpaths $\mathsf{P}_L, \mathsf{P}_R$ may be empty. If $x \in D$ or $y \in D$, then these paths are hit by $D$, and we are done. If none of $x, y$ belongs to $D$, then note that $e_i \neq e_1$ (otherwise the $P_{s+2}\ e_2 \ldots y$ implies that $y \in D$, a contradiction) and $e_i \neq e_{s+2}$ (similarly, $x$ should belong to $D$, a contradiction).

We show that, under the assumptions $x, y \notin D$ and $e_i \neq e_1, e_{s+2}$, each path $\mathsf{P}_L xy \mathsf{P}_R$ of $\mathsf{F}$ of $s + 2$ vertices is hit by $D$. Assume by contradiction that a $P_{s+2}$ path $\mathsf{P}_L xy \mathsf{P}_R$ exists which is not hit by $D$. We have three cases:

- When $\mathsf{P}_L$ and $\mathsf{P}_R$ are non-empty, since $e_i$ belongs to the $P_{s+2}\ \mathsf{P}_L xe_1 \ldots e_{s+1-|V(\mathsf{P}_L)|}$ of $\mathsf{Q}$, we deduce that $i \leq s + 1 - |V(\mathsf{P}_L)|$. Since $e_i$ belongs to the $P_{s+2}\ e_{|V(\mathsf{P}_R)|+2} \ldots e_{s+2} y \mathsf{P}_R$ of $\mathsf{Q}$, we deduce that $i \geq |V(\mathsf{P}_R)| + 2$. Then we have $|V(\mathsf{P}_R)| + 2 \leq i \leq s + 1 - |V(\mathsf{P}_L)|$, which implies $|V(\mathsf{P}_R)| + |V(\mathsf{P}_L)| \leq s - 1$, in contradiction with the assumption that $\mathsf{P}_L xy \mathsf{P}_R$ is a $P_{s+2}$.

- When $\mathsf{P}_L$ is empty, we have that $\mathsf{P}_L xy \mathsf{P}_R = xy \mathsf{P}_R$. In $\mathsf{Q}$, the path $e_{s+2} y \mathsf{P}_R$ is hit by $D'$ by hypothesis, thus either $e_{s+2} = e_i$ (a contradiction), or $\mathsf{P}_R$ is hit by $D$ (another contradiction).

- When $\mathsf{P}_R$ is empty, the reasoning is similar to the previous one.

This completes the proof of the backward direction.

To finish the proof of the claim, we show that $D$ is a minimum $s$-club set of $\mathsf{F}$ if and only if $D \cup \{e\}$ is a minimum $s$-club set of $\mathsf{Q}$. For the forward direction, by contradiction, if an $s$-club set $K$ with $|K| < |D \cup \{e\}|$ exists in $\mathsf{Q}$, then $K$ must necessarily contain a vertex from $\{e_1, \ldots, e_{s+2}\}$ (otherwise the $P_{s+2}$ induced by these vertices is not hit), and thus $K \setminus \{e\}$ is, by the affirmation proved above, an $s$-club set of $\mathsf{F}$. As $|K \setminus \{e\}| < |D|$, we have a contradiction with the hypothesis. For the backward direction, if an $s$-club set $K'$ with $|K'| < |D|$ exists in $\mathsf{F}$, then by the affirmation proved above there exists $e' \in \{e_1, \ldots, e_{s+2}\}$ such that $K' \cup \{e'\}$ is an $s$-club set of $\mathsf{Q}$. But this $s$-club set has fewer vertices that $D \cup \{e\}$, a contradiction. □

We deduce the corollary we need:

**Corollary 4.7.** *Given a graph $\mathsf{F}$ with no cycle $C_r$ such that $r \leq s + 2$, let $\mathsf{F}^+$ be the graph obtained from $\mathsf{F}$ by subdividing each edge of $\mathsf{F}$ with $s + 2$ new vertices. Then $\mathsf{F}$ has a minimum $s$-club set of size $h$ if and only if $\mathsf{F}^+$ has a minimum $s$-club set of size $h + m(\mathsf{F})$, where $m(\mathsf{F})$ is the number of edges of $\mathsf{F}$.*

Note that the girth $q^+$ of $\mathsf{F}^+$ satisfies $q^+ = q * (s + 3)$, where $q$ is the girth of $\mathsf{F}$. Then we can apply the modifications in the corollary successively $\log_{s+3} g$ times starting with the graphs $\mathsf{G}^s$, built in Section 4.1 and whose girth is at least $s + 3$, in order to obtain graphs with girth larger than $g$. Then Theorem 1 follows.

The same approach may be applied to the graphs of girth at least $s + 4$ from Corollary 4.5, to deduce that:

**Corollary 4.8.** *Let $g > 0$ be an integer. For each even integer $s \geq 2$, $s$-CLUB-VD is NP-complete for subcubic planar graphs of girth larger than $g$.*

In the next section, we show that $s$-CLUB-VD is NP-complete for subcubic planar bipartite graphs when $s$ is even, but in this case the girth is not arbitrarily large. Results for graphs with arbitrarily large girth are interesting when we want to deduce similar properties for the $k$-PVC problem (see Section 7).

**Remark 1.** In the particular case $s = 1$ (i.e. CLUSTER-VD), the NP-completeness for subcubic planar bipartite graphs is claimed in [14]. However, in our opinion the instance of CLUSTER-VD built in [14] from an instance of POSITIVE PLANAR 2-IN-3 3SAT is not always planar. An example is the instance $F = c_1 \wedge c_2 \wedge c_3 \wedge c_4$ of POSITIVE PLANAR 2-IN-3 3SAT with variable set $\{v_1, \ldots, v_6\}$ and clauses $c_1 = v_1 \vee v_4 \vee v_2, c_2 = v_3 \vee v_5 \vee v_1, c_3 = v_2 \vee v_6 \vee v_3, c_4 = v_1 \vee v_2 \vee v_3$. The graph $\Gamma$ built following the reduction in [14] has (at least) two subdivisions of $K_{3,3}$: the one with vertex set $\{v_{21}^X, v_{23}^X, v_{32}^X\} \cup \{v_{21}'^Y, v_{23}'^Y, v_{32}'^Y\}$ is due to the different (clockwise and counterclockwise) orders imposed by each of the clause gadgets $\Gamma(c_1), \Gamma(c_2)$ and $\Gamma(c_3)$ to a pair of variable gadgets among $\Gamma(v_1), \Gamma(v_2), \Gamma(v_3)$; the one with vertex set $\{v_{24}^X, v_{23}^X, v_{32}^X\} \cup \{v_{24}'^X, v_{23}'^Y, v_{32}'^Y\}$ seems merely due to the non planarity of the connections between the clause gadget $\Gamma(c_4)$ and its three corresponding variable gadgets. By Kuratowski's theorem [18] the graph $\Gamma$ is not planar.

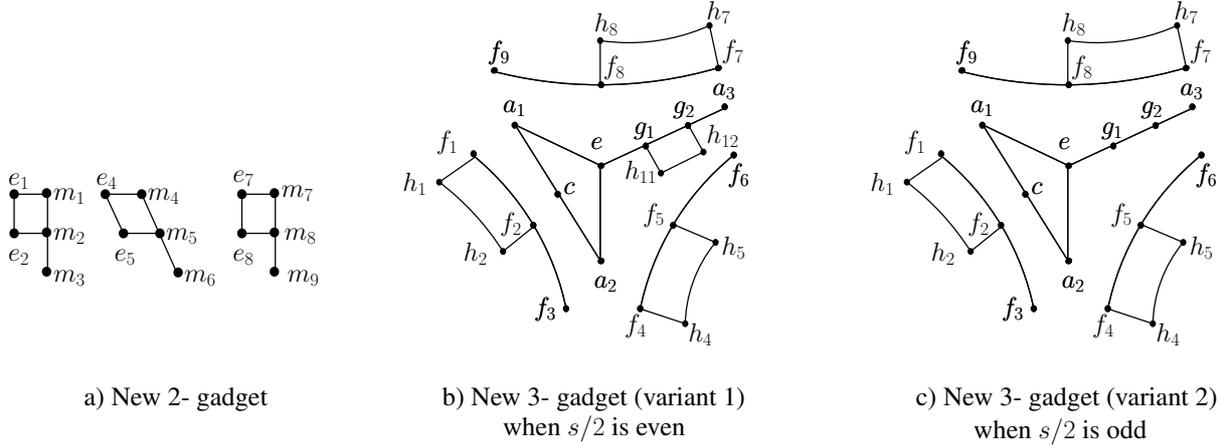

| a) New 2- gadget | b) New 3- gadget (variant 1)<br>when $s/2$ is even | c) New 3- gadget (variant 2)<br>when $s/2$ is odd |

Figure 4: New versions of the 2- and 3-gadgets.

## 5   Proof of Theorem 2

This section is devoted to the following theorem:

**Theorem 2.** *Let $s \geq 2$ be an integer. When $s$ is even, the problem $s$-CLUB-VD is NP-complete for subcubic planar bipartite graphs of girth equal to $s + 2$.*

Let $\mathsf{G}$ be the graph obtained by the construction in Section 3 and let $\mathsf{G}^s$ be the graph obtained from $\mathsf{G}$ by performing the following operations (the replacements are made using the same edges for the connections between gadgets):

- for each $T$-gadget $\mathsf{T}$ (see Figure 1b), subdivide using $s - 1$ new vertices each of the edges $xy$ such that $x \in \mathcal{E}(\mathsf{T})$ and $y \in \mathcal{C}(\mathsf{T})$.

- for each 2-gadget, replace it with the new 2-gadget in Figure 4a, then subdivide using $s - 2$ new vertices each of the edges $m_1 m_2, m_4 m_5, m_7 m_8$.

- for each 3-gadget, according to the parity of $s/2$:

  - $\star$ if $s/2$ is even, replace the 3-gadget with the new 3-gadget in Figure 4b, then subdivide using $s - 2$ new vertices each of the edges $f_1 f_2, f_4 f_5, f_7 f_8, g_1 g_2$ and subdivide with $s/2$ new vertices each of $a_1 e$ and $a_2 e$.

  - $\star$ if $s/2$ is odd, replace the 3-gadget with the new 3-gadget in Figure 4c, then subdivide using $s - 2$ new vertices each of the edges $f_1 f_2, f_4 f_5, f_7 f_8$, subdivide with $s/2$ vertices each of $a_1 e, a_2 e$ and subdivide with $s - 1$ vertices the edge $g_1 g_2$.

We keep the terminology and notations we used for an odd $s$. The approach is the same as in the proof of Theorem 1, but the paths between junction vertices connected by the chords are shorter since the subdivisions are made with $s - 2$, and not with $s - 1$ (due to the parity), vertices. The $C_4$ cycles that are now present on the chords allow to correct this flaw, but often need different local reasonings. For the seek of completeness, we give all the details of the proof in this case too.

The following results are the variants of Claim 4.1, Claim 4.2 and Corollary 4.3.

**Claim 5.1.** *The graph $\mathsf{G}^s$ is a subcubic planar bipartite graph of girth equal to $s + 2$.*

*Proof.* The planarity and maximum degree of $\mathsf{G}^s$ are immediate. The girth is given by the cycles of the chords, which are of size $s + 2$.

We prove that $\mathsf{G}^s$ is bipartite. In both cases ($s/2$ even or odd) all the gadgets are bipartite, so we consider the bipartition $(X_1, Y_1)$ of each $T$-gadget such that $\mathring{s}_w \in X_1$, the bipartition $(X_2, Y_2)$ of each 2-gadget such that $m_1, m_5, m_7 \in X_2$ and the bipartition $(X_3, Y_3)$ of each 3-gadget such that $f_1, f_4, f_7, c \in X_3$.

The graph $\mathsf{G}^s$ is bipartite with the following bipartition:

$$V_1 = \bigcup_{\substack{\text{all} \\ T-\text{gadgets in } \mathsf{G}^s}} X_1 \quad \cup \bigcup_{\substack{\text{all} \\ 2\text{-gadgets in } \mathsf{G}^s}} X_2 \quad \cup \bigcup_{\substack{\text{all} \\ 3\text{-gadgets in } \mathsf{G}^s}} X_3$$

$$V_2 = \bigcup_{\substack{\text{all} \\ T-\text{gadgets in } \mathsf{G}^s}} Y_1 \quad \cup \bigcup_{\substack{\text{all} \\ 2\text{-gadgets in } \mathsf{G}^s}} Y_2 \quad \cup \bigcup_{\substack{\text{all} \\ 3\text{-gadgets in } \mathsf{G}^s}} Y_3.$$

$\square$

**Claim 5.2.** *The following affirmations hold:*

*(a) Each $T$-gadget $\mathsf{T}$ admits exactly $s+2$ minimum $s$-club sets: $\mathcal{C}(\mathsf{T}), \mathcal{D}(\mathsf{T}), \mathcal{E}_0(\mathsf{T}), \mathcal{E}_1(\mathsf{T}), \ldots, \mathcal{E}_{s-1}(\mathsf{T})$. Their size is 12.*

*(b) The minimum $s$-club sets of a 2-gadget are of size 3.*

*(c) The minimum $s$-club sets of a 3-gadget are of size 6.*

*Proof.* Affirmation (a) is immediate, due to the distance of $s+2$ between two consecutive vertices of the same color in each $T$-gadget of $\mathsf{G}^s$. In a 2- or 3-gadget, each chord has exactly two $P_{s+2}$, each of them containing the middle standard vertex of the chord. Thus each chord has an $s$-club set of size 1, implying (b) in case of a 2-gadget and, in case of a 3-gadget, contributing a total of three vertices (for the three chords) to an $s$-club set of the 3-gadget. It remains to show that the star of the 3-gadget (variant 1 or 2) has a minimum $s$-club set of 3. The cycle with vertices $a_1, SV(a_1, e), e, SV(a_2, e), a_2, c$ is of size $2 \cdot s/2 + 4$, that is $s + 4$, so it needs to be hit by two vertices of $D$. The path induced by the vertices $h_{11}, g_1, SV(g_1, g_2), g_2, a_3$ has $s + 2$ vertices in variant 1 of the 3-gadget, and so is the path joining $g_1$ to $a_3$ in variant 2 of the 3-gadget. So at least three vertices of $D$ are needed to hit all the $P_{s+2}$ in the star of the 3-gadget. Since $\{a_1, a_2, g_1\}$ is an $s$-club set of the star, affirmation (c) follows. $\square$

The corollary is easily deduced:

**Corollary 5.3.** *Each $s$-club set of $\mathsf{G}$ has size at least $12|T| + 3n_2 + 6n_3$.*

Again, the following result is similar to Claim 4.4, but with many different details in the proof.

**Claim 5.4.** *Let $s$ be an even integer, with $s \geq 2$. There exists a 3D-matching $M \subseteq T$ if and only if $\mathsf{G}^s$ contains an $s$-club set of size equal to $12|T| + 3n_2 + 6n_3$.*

*Proof.* The $\Rightarrow$ direction is proved similarly to the case where $(s-1)/2$ is even in Claim 4.4, *i.e.* using only $s$-club sets of size 6 in the 3-gadgets (defined identically).

The $\Leftarrow$ direction also follows the lines in the proof of Claim 4.4, but some results need deeper arguments in the current case. Let $D$ be an $s$-club set of $\mathsf{G}^s$ of size $12|T| + 3n_2 + 6n_3$. Recall that $SV(g_1, g_2)$ is of size $s - 2$ when $s/2$ is even and $s - 1$ otherwise. Then $D$ has the following properties:

(a) $D$ hits:

($a_1$) each $T$-gadget by exactly 12 vertices, those of a unique color among $\mathcal{C}, \mathcal{D}$ and $\mathcal{E}_q$ ($0 \leq q \leq s - 1$).

($a_2$) each chord of a 2- or 3-gagdget by exactly one vertex.

($a_3$) the star of each 3-gadget by exactly 3 vertices.

*Proof.* Follows by Claim 5.2 and Corollary 5.3, since $D$ has the minimum possible size.

(b) In a 3-gadget, we cannot have simultaneously $a_1, a_2, a_3 \in D$.

*Proof.* In the contrary case, since $D$ can contain only three vertices of the star, we deduce that the path joining $e$ to $h_{12}$ through the subdivision vertices when $s/2$ is even (respectively the path joining $e$ to $g_2$ through the subdivision vertices when $s/2$ is odd) is a $P_{s+2}$ that is not hit, a contradiction.

(c) Each junction vertex $v$ not belonging to $D$ is the endpoint, in the $T$-gadget $\mathsf{T}$ containing it, of a path non-hit by $D$ which is:

($c_1$) a $P_{s+1}$ if $\mathsf{T}$ is either $\mathcal{C}$- or $\mathcal{D}$-colored by $D$.

($c_2$) a $P_h$ with $h \geq s/2 + 1$ when $\mathsf{T}$ is $\mathcal{E}_q$-colored by $D$ ($0 \leq q \leq s - 1$).

The vertex set of this path is denoted by $L(v)$.

*Proof.* As in Claim 4.4, since two consecutive vertices of $V(\mathsf{T}) \cap D$ are at distance $s + 2$, the $s + 1$ vertices between them allow us to build two paths non-hit by $D$ along $\mathsf{T}$, one in each direction (clowkwise, counterclockwise). Under the hypothesis of affirmation ($c_1$), the conclusion follows from the observation that one of these paths is empty and the other one has maximum length. Under the hypothesis of affirmation ($c_2$), the longest of the two paths is as short as possible when the two paths have equal length. More precisely, they have $s/2 + 1$ vertices each when $q = s/2$ (if $v \in \mathcal{D}$) respectively when $q = s/2 - 1$ (if $v \in \mathcal{C}$).

(d) For each junction vertex $v$ from a $T$-gadget $\mathsf{T}$, the set $D$ contains:

($d_1$) at least one vertex from the set $\{v, J(v)\}$, when $\mathsf{T}$ is either $\mathcal{C}$- or $\mathcal{D}$-colored by $D$.

($d_2$) at least one vertex from each $P_{s/2+1}$ with endpoint $J(v)$ and the other vertices in the 2- or 3-gadget containing $J(v)$, when $\mathsf{T}$ is $\mathcal{E}_q$-colored by $D$.

*Proof.* Affirmation ($d_1$) immediately follows from affirmation ($c_1$) since the path induced by $L(v) \cup \{J(v)\}$ is a $P_{s+2}$ that must be hit by $D$. Affirmation ($d_2$) follows from the remark that the path induced by $L(v)$, according to affirmation ($c_2$), and the $s/2 + 1$ vertices of the path starting with $J(v)$ has at least $s+2$ vertices, and thus it must be hit by $D$.

From these properties, we deduce following the steps below that exactly one of the $T$-gadgets connected with a 2- or 3-gadget is $\mathcal{C}$-colored, whereas the other $T$- gadgets are $\mathcal{D}$-colored.

(e) When two $T$-gadgets are connected by the same 2- or 3-gadget, at least one of them is $\mathcal{D}$-colored by $D$.

*Proof.* Assume by contradiction that two $T$-gadgets exist neither of which is $\mathcal{D}$-colored and such that they are both connected to the same 2- or 3-gadget. Without loss of generality and in order to fix ideas, we assume the two gadgets are denoted by $\mathsf{T}_i$ and $\mathsf{T}_j$ and we consider the chord (say with vertices $m_1, m_2, m_3, e_1, e_2$ w.l.o.g.) connecting $\overset{\bullet}{s}^i_w$ to $\overset{\bullet}{x}^j_w$. By affirmation ($c_2$), the path with vertex set given by $L(\overset{\bullet}{s}^i_w), m_1, SV(m_1, m_2), m_2, m_3$ and $L(\overset{\bullet}{x}^j_w)$ has at least $2(s/2 + 1) + (s - 2) + 3 = 2s + 3$ vertices. Since the chord must be hit by exactly one vertex, the path has exactly $2s + 3$ vertices and its central vertex $v$ must belong to $D$. Then $v$ is the $s/2$-th subdivision vertex when these vertices are ranked by their distance to $m_1$, except in the case $s = 2$ when $v = m_2$. Then, when $s \neq 2$, the path containing $s/2 - 1$ vertices preceding $v$ in this ranking, $m_1, e_1, e_2, m_2, m_3$ and $L(\overset{\bullet}{x}^j_w)$ has at least $s/2 - 1 + 5 + s/2 + 1 = s + 5$ vertices and is not hit by $D$, a contradiction. When $s = 2$, the path $e_2 e_1 m_1 \overset{\bullet}{s}^i_w$ is a $P_{s+2}$ that is not hit by $D$, a contradiction. Affirmation (e) follows.

(f) No $T$-gadget is $\mathcal{E}_q$-colored by $D$ ($0 \leq q \leq s - 1$).

*Proof.* We assume by contradiction that a $T$-gadget is $\mathcal{E}_q$-colored by $D$, and study the different cases.

- If a 2-gadget is connected with a $T$-gadget $\mathsf{T}_i$ that is $\mathcal{E}_q$-colored by $D$, we deduce by affirmation (e) that $\mathsf{T}_j$ is $\mathcal{D}$-colored by $D$ and by affirmation ($d_1$) that $m_6 \in D$. But then affirmation ($d_2$) is contradicted by $v = \overset{\circ}{u}^i_w$ since the path induced by $\{m_4, m_5\} \cup SV(m_4, m_5)$ has length $s$ (which is at least equal to $s/2 + 1$ and is nit hit by $D$. (The reasoning is similar when $\mathsf{T}_j$, instead of $\mathsf{T}_i$, is $\mathcal{E}_q$-colored by $D$.)

- If a 3-gadget is connected with a $T$-gadget $\mathsf{T}_i$ that is $\mathcal{E}_q$-colored by $D$, we deduce by affirmation (e) that $\mathsf{T}_j, \mathsf{T}_k$ are $\mathcal{D}$-colored by $D$, by affirmation ($d_1$) that $a_2, a_3 \in D$ and by affirmation (b) that $a_1 \notin D$. By affirmation ($d_2$) applied to the junction vertex $\overset{\circ}{u}^i_w$, $D$ contains one of the vertices in $SV(a_1, e)$. Note that now $D$ contains three vertices of the star of the 3-gadget. By affirmation ($a_3$) this value cannot be exceeded. Then the path with vertices $h_{12}, g_2, SV(g_1, g_2), g_1, e$ when $s/2$ is even (respectively the path with vertices $g_2, SV(g_1, g_2), g_1, e$ when $s/2$ is odd) has $s + 2$ vertices and is not hit by $D$, a contradiction.

- The reasoning is similar in the case where $\mathsf{T}_j$, instead of $\mathsf{T}_i$, is $\mathcal{E}_q$-colored by $D$, and only slightly different when $\mathsf{T}_k$, instead of $\mathsf{T}_i$, is $\mathcal{E}_q$-colored by $D$. In the latter case, we similarly deduce that $a_1, a_2 \in D$ and $a_3 \notin D$. The third, and last, vertex of $D$ on the star must belong to the path joining $a_3$ to $e$. The path with vertices given by $L(\mathring{\mathring{u}}_w^k)$, $a_3$, $g_2$, $SV(g_1, g_2)$, $g_1$, $e$ and $SV(a_1, e)$ has at least $(s/2 + 1) + 4 + (s - 2) + s/2 = 2s + 3$ vertices when $s/2$ is even, and at least $2s + 4$ vertices when $s/2$ is odd. In the latter case, two vertices of $D$ are needed to hit all the $P_{s+2}$ subpaths of the path, but only one is available, a contradiction. So $s/2$ is even, and the above-mentioned path must have exactly $2s + 3$ vertices in order to allow all its $P_{s+2}$ subpaths to be hit by exactly one vertex $v$. Then $v$ is necessarily the center of the path, that is, it is the $(s/2 - 1)$-th subdivision vertex of $SV(g_1, g_2)$ when going from $g_2$ to $g_1$. Then, with the notation $SV'$ for the set of $s/2 - 2$ subdivision vertices between $v$ and $g_2$, the path induced by $SV' \cup \{g_2, h_{12}, h_{11}, g_1, e\} \cup SV(e, a_1)$ has $s/2 - 2 + 5 + s/2 = s + 3$ vertices and is not hit by $D$, a contradiction.

(g) Each 2- or 3-gadget is connected with at most one $T$-gadget $\mathcal{C}$-colored by $D$ (similar proof to that in Claim 4.4).

(h) No 2-gadget is connected with two $T$-gadgets $\mathcal{D}$-colored by $D$ (similar proof to that in Claim 4.4).

(i) No 3-gadget is connected with three $T$-gadgets $\mathcal{D}$-colored by $D$ (similar proof to that in Claim 4.4).

We deduce that each 2- and 3-gadget is connected with exactly one $T$-gadget $\mathcal{C}$-colored by $D$, and we define $M$ to be the set containing all the $T$-gadgets that are $\mathcal{C}$-colored by $D$. □

Since $\mathsf{G}^s$ is subcubic, planar and bipartite according to Claim 5.1, and since $s$-CLUB-VD belongs to NP, Theorem 2 follows for graphs with girth equal to $s + 2$.

It remains to turn subcubic instances into cubic instances, and this is done in the next section.

# 6   Proof of Theorem 3

The theorem we prove in this section is the following one:

**Theorem 3.** *Let $s \geq 1$ be an integer. The problem $s$-CLUB-VD ($s \geq 1$) is NP-complete for cubic planar bipartite graphs.*

We show how to appropriately transform subcubic instances into cubic instances. The main idea is to make disjoint pairs and triplets of 2-degree vertices from $\mathsf{G}^s$ ($s \geq 1$), and to connect each pair or triplet to a new gadget all of whose remaining vertices are already of degree 3.

## 6.1   Tools

Let $s \geq 1$ be an integer. Let $\mathsf{A}_s$ and $\mathsf{B}_s$ be the gadgets in Figure 5 when $s$ is odd, and the gadgets in Figure 6 when $s$ is even. The operation of identifying two vertices $x, y$ of a graph $\mathsf{F}$ (respectively three vertices $x, y, z$ of $\mathsf{F}$) with the vertices $a, b$ of $\mathsf{A}_s$ ($a, b, c$ of $\mathsf{B}_s$) is called a 2-*completion* of $x, y$ (a 3-*completion* of $x, y, z$). Note that $\mathsf{A}_s$ is a stand-alone graph, whereas $\mathsf{B}_s$ is computed using 2-completions (represented by grey squares in Figure 5c and Figure 6b). Each 2-completion uses a distinct copy of $\mathsf{A}_s$.

**Claim 6.1.** *The following affirmations hold:*

(a) $\mathsf{A}_s$ *is a planar bipartite graph, whose vertices except $a, b$ have degree 3, and in which $a, b$ belong to different sides of the bipartition.*

(b) $\mathsf{B}_s$ *is a planar bipartite graph, whose vertices except $a, b, c$ have degree 3, and in which $a, b, c$ belong to the same side of the bipartition.*

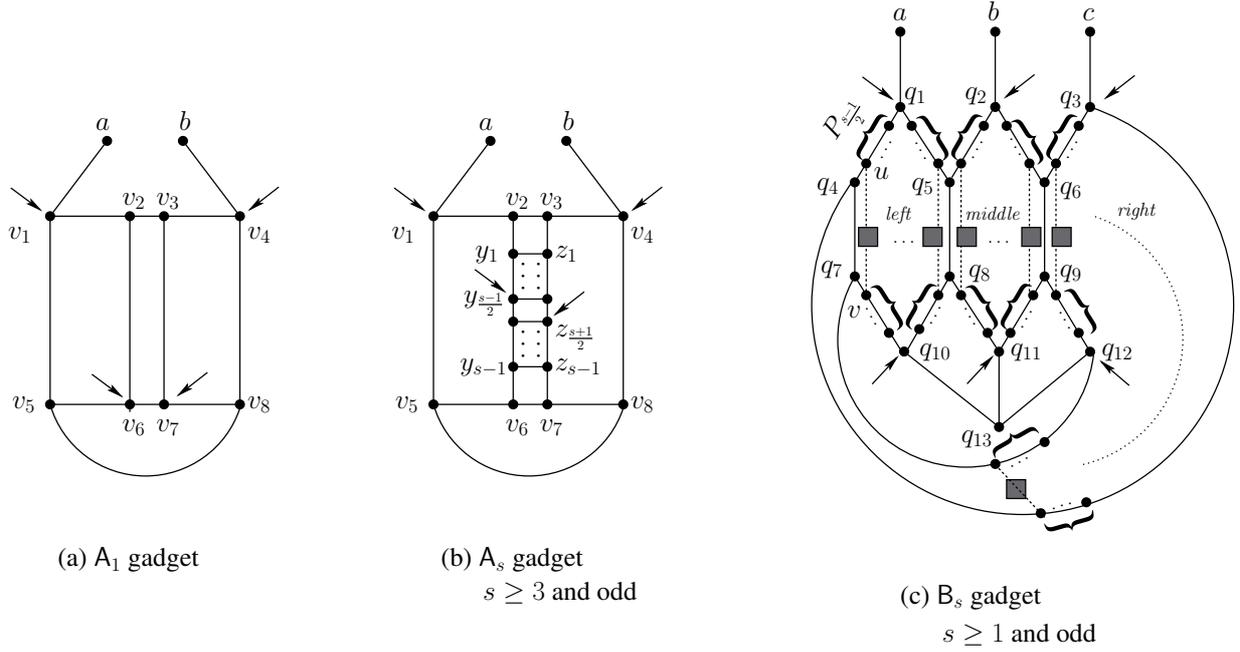

(a) $\mathsf{A}_1$ gadget

(b) $\mathsf{A}_s$ gadget
$s \geq 3$ and odd

(c) $\mathsf{B}_s$ gadget
$s \geq 1$ and odd

Figure 5: Gadgets $\mathsf{A}_s$ and $\mathsf{B}_s$ for odd $s$, $s \geq 1$. The arrows show a minimum $s$-club set. (a) $\mathsf{A}_1$. (b) In $\mathsf{A}_s$ ($s \geq 3$ and odd), all the edges $y_c z_c$ exist, for $1 \leq c \leq s - 1$. (c) In $\mathsf{B}_s$, the brackets indicate $P_{(s-1)/2}$ paths. Each vertex $u$ of such a path has a symmetric $v$ (see the example of $u$ on the path between $q_4$ and $q_1$), and the pair $(u, v)$ undergoes a 2-completion (indicated by the dotted line with a grey square on it). Only 6 2-completions are represented on the figure, two for each of the three cycles identified as left, middle and right. There are $2 \cdot (s - 1)/2$ such completions on each of the these cycles.

*Proof.* In $\mathsf{A}_s$, all the cycles are even, so the graph is bipartite. Figure 5a,b and Figure 6a show embeddings of the graph in the plane, when $s$ is odd and even respectively. The degree equal to 3 for all the vertices but $a, b$ and the odd-length paths between $a$ and $b$ are easily checked on the figures. Affirmation (a) is proved.

In $\mathsf{B}_s$, when the 2-completions are left apart, the bipartiteness and planarity are easily checked. The vertices $q_h$, for $1 \leq h \leq 13$ have degree 3 whereas the subdivision vertices have (initial) degree 2, but the 2-completions add one neighbor to each of them. The 2-completions are performed for pairs of vertices that avoid the intersections in the plane, so the resulting graph is planar.

Moreover, each pair $(u, v)$ of vertices of $\mathsf{B}_s$ that undergoes a 2-completion has been defined so as to satisfy $u \in Y_1, v \in Y_2$, where $(Y_1, Y_2)$ is the bipartition of $V(\mathsf{B}_s)$ when 2-completions are left apart. In $\mathsf{A}_s$, assume $a \in X_1, b \in X_2$, where $(X_1, X_2)$ is the bipartition of $V(\mathsf{A}_s)$. When $u, v$ from $\mathsf{B}_s$ are identified with $a, b$ respectively from $\mathsf{A}_s$, we have that $(X_1 \setminus \{a\} \cup Y_1, X_2 \setminus \{b\} \cup Y_2)$ is a bipartition of $V(\mathsf{B}_s)$ when the 2-completion of $(u, v)$ has been performed, and the other 2-completions are left apart. Repeating this reasoning for all the 2-completions we deduce that $\mathsf{B}_s$ is bipartite. In this bipartition, $a, b, c$ are on the same side since all the paths joining any two of them are even, and affirmation (b) follows. □

The two following claims indicate how 2- and 3-completions affect the minimum $s$-club set of a graph.

**Claim 6.2.** *Let $s \geq 1$ and let $\mathsf{Q}$ be the graph obtained from a graph $\mathsf{F}$ by performing a 2-completion. Then $\mathsf{F}$ has a minimum $s$-club of size $h$ if and only if $\mathsf{Q}$ has a minimum $s$-club set of size $h + 4$.*

*Proof.* Let $r = s - 1$ when $s$ is odd, and $r = s$ when $s$ is even. Then $s + 2 = r + 3$ when $s$ is odd and $s + 2 = r + 2$ when $s$ is even. The cycles $v_1 v_2 y_1 \ldots y_r v_6 v_5 v_1$ and $v_3 v_4 v_8 v_7 z_r \ldots z_1 v_3$ of $\mathsf{A}_s$ (see Figure 5a,b and Figure 6a) are disjoint and have $r + 4$ vertices each. Thus an $s$-club set of $\mathsf{A}_s$ must hit each of these cycles using at least two vertices, for a total of at least 4 vertices. We deduce that the set $D'$ of size 4 indicated in Figure 5a,b and Figure 6a reaches the minimum possible cardinality for an $s$-club set. To see that $D'$ is an $s$-club set, note that in the graph induced by $V(\mathsf{A}_s) \setminus D'$ there are two connected components, respectively containing $v_2$ and $v_5$, each of whose has size at most $s + 2$. However, none of them is a $P_{s+2}$.

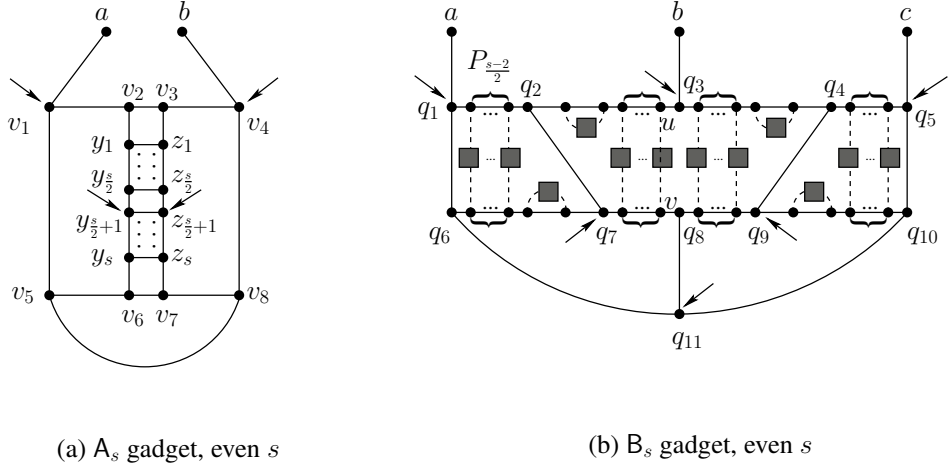

(a) $\mathsf{A}_s$ gadget, even $s$          (b) $\mathsf{B}_s$ gadget, even $s$

Figure 6: Gadgets $\mathsf{A}_s$ and $\mathsf{B}_s$ for even $s$, $s \geq 2$. The arrows show a minimum $s$-club set. (a) In $\mathsf{A}_s$ all the edges $y_c z_c$ exist, for $1 \leq c \leq s$. (b) In $\mathsf{B}_s$, the brackets indicate $P_{(s-2)/2}$ paths. Each vertex $u$ of such a path has a symmetric $v$, and the pair $(u, v)$ undergoes a 2-completion (indicated by the dotted line with a grey square on it). See the example of the vertex $u$ (left neighbor of $q_3$). Only 8 such 2-completions are represented on the figure, but there are $4 \cdot (s-2)/2$ such 2-completions in $\mathsf{B}_s$. Moreover, four other pairs of vertices undergo a 2-completion, for a total of $2s$ 2-completions.

Thus $D'$ is a minimum $s$-club set of $\mathsf{A}_s$, and has cardinality 4. Given a minimum $s$-club set $D$ of $\mathsf{F}$ of size $h$, the set $D \cup D'$ is therefore a minimum $s$-club set of $\mathsf{Q}$ of size $h + 4$, since the vertices $v_1$ and $v_4$ of $D'$ also belong to all the $P_{s+2}$ with vertices from both $V(\mathsf{F})$ and $V(\mathsf{A}_s) \setminus \{a, b\}$. The converse is immediate. □

**Claim 6.3.** *Let $s \geq 1$ be an integer, and let $\mathsf{Q}$ be the graph obtained from a graph $\mathsf{F}$ by performing a 3-completion. Then $\mathsf{F}$ has a minimum $s$-club set of size $h$ if and only if $\mathsf{Q}$ has a minimum $s$-club set of size $h + 12s - 6$ (when $s$ is odd) or $h + 8s + 6$ (when $s$ is even).*

*Proof.* We first show that a minimum $s$-club set of $\mathsf{B}_s$, when the 2-completions are left apart, is 6. Then we count the number of 2-completions and conclude.

**When $s$ is odd.** Consider the cycles $q_4 \ldots q_1 \ldots q_5 \ldots q_2 \ldots q_6 \ldots q_3 \ldots q_4$ (called *top* cycle) and $q_7 \ldots q_{10} \ldots q_8 \ldots q_{11} \ldots q_9 \ldots q_{12} \ldots q_7$ (called *bottom* cycle), where the dots indicate a $P_{(s-1)/2}$ represented by a bracket in Figure 5b. Each of them has $6(s-1)/2 + 6 = 3s + 3$ vertices, and they are disjoint. When $s > 1$, we have $3s + 3 > 2s + 4$ and thus an $s$-club set must hit each of these cycles using at least 3 vertices, for a total of at least 6. The set $D'$ indicated by the arrows on the figure is then an $s$-club set of minimum size of $\mathsf{B}_s$.

We focus now on the case $s = 1$. The paths $P_{(s-1)/2}$ are empty, and the cycles $q_1 q_4 q_7 q_{10} q_8 q_5 q_1$, $q_2 q_5$ $q_8 q_{11} q_9 q_6 q_2$ and $q_3 q_6 q_9 q_{12}$ $q_7 q_4 q_3$ are $C_6$ cycles. Each of them shares two vertices and one edge with the two other cycles, implying that there is a symmetry between them (despite the drawing). The set $D' = \{q_1, q_2, q_3, q_{10}, q_{11}, q_{12}\}$ indicated by arrows on the figure is a 1-club set of $\mathsf{B}_1$. We show that any 1-club set $D_0$ of $\mathsf{B}_1$ has at least 6 vertices.

If each of the top and bottom cycles is hit by $D_0$ using at least three vertices, then the conclusion follows. Otherwise, assume the top cycle is hit by $D_0$ using at most two vertices (the reasoning is similar for the bottom cycle). Without loss of generality we may assume that $q_1, q_6 \in D_0$, since all the other possibilities are symmetric with this one. The following $P_3$ are disjoint, thus need to be hit by at least one vertex each: $q_3 q_4 q_7$, $q_9 q_{12} q_{13}$, $q_{11} q_8 q_{10}$. When each of them is hit by exactly one vertex, we have $|D_0| = 5$, otherwise we are done. In the former configuration, we have two cases. If $q_8 \notin D_0$, then the $P_3$ $q_2 q_5 q_8$ adds a new vertex to $D_0$ and we are done. If $q_8 \in D_0$, and given that $q_{11} q_8 q_{10}$ is hit exactly once, we deduce $q_{10}, q_{11} \notin D_0$ and the $P_3$ $q_{11} q_{13} q_{10}$ implies $q_{13} \in D_0$. The $P_3$ $q_9 q_{12} q_{13}$ is also hit exactly once, thus $q_9, q_{12} \notin D_0$. But then $q_{12} q_9 q_{11}$ is a $P_3$ that is not hit by $D_0$, a contradiction.

We deduce that for an odd $s$, the $s$-club set $D'$ shown in Figure 5c together with the appropriate $s$-club sets of the $\mathsf{A}_s$ gadgets (according to Claim 6.2) is a minimum $s$-club set of $\mathsf{B}_s$. There are $6(s-1)/2 = 3s - 3$ 2-completions, and by Claim 6.2 each of them adds 4 vertices to the minimum $s$-club set $D'$ of $\mathsf{B}_s$, whose cardinality is thus $6 + 4(3s - 3) = 12s - 6$. Since $D'$ contains the neighbors of $a, b, c$, it will also hit all the $P_{s+2}$-paths with vertices from both $\mathsf{F}$ and $\mathsf{B}_s$. Then $D'$ may be added to any minimum $s$-club set of the initial graph $\mathsf{F}$ (that undergone a 3-completion) to obtain a minimum $s$-club set of the resulting graph $\mathsf{Q}$. The other direction of the claim is immediate.

**When $s$ is even.** Let $D'$ be a minimum $s$-club of $\mathsf{B}_s$ when the 2-completions are left apart. We want to show that $|D'| \geq 6$. We have two cases, depending on the number of vertices of $D'$ hitting the two bottom cycles of length $s+4$ and containing both $q_8$ and $q_{11}$ (see Figure 6b). We note that the two bottom cycles cannot be hit altogether by only two vertices, since then the two vertices would necessarily be $q_8$ and $q_{11}$. But they do not hit all the $P_{s+2}$ in the two bottom cycles.

When the two bottom cycles are hit by two distinct vertices each (for a total of four vertices), we deduce that $|D'| \geq 6$ by noticing that the path with $2s+5$ vertices (on the top of the figure) joining $q_1$ to $q_5$ through $q_2, q_3$ and $q_4$ is also hit by at least two vertices. When the two bottom cycles are hit altogether by three vertices, that means either $q_8 \in D'$ or $q_{11} \in D'$, and each of the horizontal paths joining $q_6$ to $q_8$ (through $q_7$) and $q_8$ to $q_{10}$ (through $q_9$) is hit by exactly one vertex of $D$. Each of the following cycles and paths needs to be hit by at least one *additional* vertex of $D'$: the $C_{s+4}$ cycle $q_1 ... q_2 q_7 ... q_6 q_1$, the $P_{s+3}$ induced by $q_3$ together with the $s/2+1$ vertices to its left and the $s/2+1$ vertices to its right, the $C_{s+4}$ cycle $q_4 ... q_5 q_{10} ... q_9 q_4$. Then $|D'| \geq 6$.

The set $D'$ defined in Figure 6b by the arrows is then a minimum $s$-club set of $\mathsf{B}_s$ when the 2-completions are left apart. There are $2s$ 2-completions, each of which adds 4 vertices to the minimum $s$-club set $D'$ of $\mathsf{B}_s$. The conclusion follows. □

## 6.2   The proof

In order to prove Theorem 3, we now reduce $s$-CLUB-VD ($s \geq 1$) for the subcubic planar bipartite instances $\mathsf{G}^s$ previously built to $s$-CLUB-VD for cubic planar bipartite instances. By Claims 6.2 and 6.3, it is sufficient to show that we can partition the set of 2-degree vertices of $\mathsf{G}^s$ into pairs and triplets of vertices for which a 2- or a 3-completion respectively can be performed such that the resulting graph is planar and bipartite. The former condition needs to carefully choose the pairs and triplets on the same facet of the planar embedding, avoiding intersections. The latter condition is satisfied using Claim 6.1: we choose pairs of vertices on different sides of the bipartition of $\mathsf{G}^s$ and triplets of vertices on the same side of the bipartition of $\mathsf{G}^s$.

**When $s$ is odd.** We define the following pairs and triplets in the graph $\mathsf{G}^s$ defined in Section 4:

- for each $T$-gadget $\mathsf{T}$ and each support $\mathsf{Supp}_\mathsf{T}(w)$ of it: $(\mathring{s}_w, \dot{s}_w, \overset{\bullet}{u}_w)$, $(\dot{u}_w, \mathring{x}_w)$, $(\dot{x}_w, \mathring{z}_w)$, $(\overset{\bullet}{z}_w, \dot{z}_w)$ and, for each set of $s-1$ consecutive subdivision vertices, define $(s-1)/2$ disjoint pairs of consecutive subdivision vertices. There are four such sets for each $w$, namely $SV(\dot{s}_w, \mathring{u}_w)$, $SV(\dot{u}_w, \mathring{x}_w)$, $SV(\dot{x}_w, \mathring{z}_w)$ and $SV(\dot{z}_w, \mathring{s}_{w'})$ where $w'$ is the element of $W$ whose support comes next along $\mathsf{T}$ in counterclockwise direction.

- for each 2-gadget: $(m_1, m_2), (m_3, m_4, m_6), (m_5, m_7), (m_8, m_9)$ and, for each set of $s-1$ consecutive subdivision vertices, define $(s-1)/2$ disjoint pairs of consecutive subdivision vertices. There are three such sets for each 2-gadget, namely $SV(m_1, m_2)$, $SV(m_4, m_5)$ and $SV(m_7, m_8)$.

- for each 3-gadget: $(f_1, f_2), (f_4, f_5), (f_7, f_8), (f_3, c), (a_3, f_6, g_1), (g_2, f_9)$, and, for each of the following sets, form disjoint pairs of consecutive subdivision vertices: $SV(f_1, f_2), SV(f_4, f_5), SV(f_7, f_8), SV(g_1, g_2)$, $SV(a_1, e)$ and $SV(a_2, e)$. The four first sets give $(s-1)/2$ pairs each, whereas the two last sets yield $(s-1)/4$ ( or $(s-1)/4 + (s+2)/2$) pairs each when $(s-1)/2$ is even (odd, respectively).

**When $s$ is even.** We define the following pairs and triplets in the graph $\mathsf{G}^s$ defined in Section 5:

- for each $T$-gadget $\mathsf{T}$ and each support $\mathsf{Supp}_\mathsf{T}(w)$ of it: $(\mathring{s}_w, v)$, where $v$ is the neighbor of $\dot{s}_w$ that is a subdivision vertex, $(\overset{\bullet}{u}_w, \dot{u}_w), (\mathring{z}_w, \overset{\bullet}{z}_w)$ and, for each of the following sets, form disjoint pairs of consecutive subdivision vertices: $SV(\dot{s}_w, \mathring{u}_w) \setminus \{v\}$, $SV(\dot{u}_w, \mathring{x}_w) \cup \{\overset{\bullet}{x}_w\}$, $\{\dot{x}_w\} \cup SV(\dot{x}_w, \mathring{z}_w)$ and $\{\dot{z}_w\} \cup SV(\dot{z}_w, \mathring{s}_{w'})$ where $w'$ is the element of $W$ whose support comes next along $\mathsf{T}$ in counterclockwise direction. These sets give $(s-2)/2, s/2, s/2$ and $s/2$ pairs respectively. Note that $\dot{s}_w$ has not been used yet.

- for each 2-gadget:
  - ★ if $s \geq 4$: $(e_1, e_2)$, $(\mathring{s}^i_w, m_3, v_1)$ where $v_1$ is the subdivision vertex adjacent to $m_4$, $(e_4, v_2)$ where $v_2$ is the subdivision vertex adjacent to $v_1$, $(e_5, m_6, v_3)$ where $v_3$ is the subdivision vertex adjacent to $m_7$, $(v_4, m_9, \mathring{s}^j_w)$ where $v_4$ is the subdivision vertex adjacent to $v_3$, $(e_7, e_8)$ and, for each of the following sets, form disjoint pairs of consecutive subdivision vertices: $SV(m_1, m_2)$, $SV(m_4, m_5) \setminus \{v_1, v_2\}$, $SV(m_7, m_8) \setminus \{v_3, v_4\}$. These sets give $(s-2)/2, (s-4)/2, (s-4)/2$ pairs respectively. To ensure the planarity of the drawings, $e_4, e_5$ ($e_7, e_8$) must be drawn to the right of $m_4, m_5$ ($m_7, m_8$ respectively).
  - ★ if $s = 2$, modify first the pairs of $\mathsf{Supp}_{\mathsf{T}_i}(w)$ in order to replace the pairs $(\mathring{s}^i_w, v)$, where $\{v\} = SV(\dot{s}^i_w, \mathring{u}^i_w)$, $(\dot{u}^i_w, \dot{u}^i_w)$ and $(v', \mathring{x}^i_w)$, where $\{v'\} = SV(\dot{u}^i_w, \mathring{x}^i_w)$ with the triplet $(\mathring{s}^i_w, \dot{s}^i_w, \mathring{x}^i_w)$ and the pair $(\dot{u}^i_w, v')$. Then $v$ and $\dot{u}^i_w$ are not yet used in any pair or triplet. In $\mathsf{Supp}_{\mathsf{T}_j}(w)$, only $\dot{s}^j_w$ is not yet used. Then we define the following pairs and triplets: $(e_1, e_2)$, $(v, e_4)$, $(m_3, e_5)$, $(\dot{u}^i_w, e_7, m_6)$, $(e_8, m_9, \dot{s}^j_w)$. To ensure the planarity of the drawings, $e_4, e_5$ ($e_7, e_8$) must be drawn to the left of $m_4, m_5$ ($m_7, m_8$ respectively).
- for each 3-gadget:
  - ★ if $s/2$ is even: $(h_1, h_2)$, $(h_4, h_5)$, $(h_7, h_8)$, $(\dot{s}^i_w, f_3, c)$, $(\dot{s}^j_w, f_6, v_2)$ where $v_2$ is the subdivision vertex adjacent with $a_2$, $(v_1, v_3)$ where $v_1, v_3$ are the subdivision vertices adjacent with $a_1$ and $v_2$ respectively, $(f_9, v_4)$ where $v_4$ is the subdivision vertex adjacent with $v_1$, $(a_3, \dot{s}^k_w)$ (note that $s \geq 4$ in this case, so all these vertices exist) and, for each of the following sets, form disjoint pairs of consecutive subdivision vertices: $SV(f_1, f_2), SV(f_4, f_5), SV(f_7, f_8), SV(a_2, e) \setminus \{v_2, v_3\}, SV(a_1, e) \setminus \{v_1, v_4\}, SV(g_1, g_2)$. These sets give respectively $(s-2)/2, (s-2)/2, (s-2)/2, (s/2-2)/2, (s/2-2)/2, (s-2)/2$ pairs.
  - ★ if $s/2$ is odd: $(h_1, h_2)$, $(h_4, h_5)$, $(h_7, h_8)$, $(h_{11}, h_{12})$, $(\dot{s}^i_w, f_3, c)$, $(\dot{s}^j_w, f_6, v_2)$ where $v_2$ is the subdivision vertex adjacent to $a_2$, $(\dot{s}^k_w, f_9, v_1)$ where $v_1$ is the subdivision vertex adjacent to $a_1$, $(a_3, g_2)$ and, for each of the following sets, form disjoint pairs of consecutive subdivision vertices: $SV(f_1, f_2), SV(f_4, f_5), SV(f_7, f_8), SV(a_2, e) \setminus \{v_2\}, SV(a_1, e) \setminus \{v_1\}, \{g_1\} \cup SV(g_1, g_2)$. These sets give respectively $(s-2)/2, (s-2)/2, (s-2)/2, (s/2-1)/2, (s/2-1)/2, s/2$ pairs.

In each case, denote by $k_2(s)$ the total number of pairs and by $k_3(s)$ the total number of triplets. Let $\mathsf{G}^s$ be the graph obtained from $\mathsf{G}^s$ by performing the 2- and 3-completions of these pairs and triplets. For each of them, an arbitrary bijection is chosen between the vertices of each pair (triplet) and the vertices $a, b$ ($a, b, c$ respectively) of $\mathsf{A}_s$ ($\mathsf{B}_s$ respectively). The pairs and triplets we chose are not intersecting, in the sense where one may draw lines between each two vertices from a pair or triplet in a way that no two lines intersect (except in their endpoints). Then Claim 4.1, Claim 5.1 and Claim 6.1 ensure that the resulting graph is planar. Moreover, the parity of the paths between two vertices of a pair or of a triplet is the same in $\mathsf{G}^s$ and in the gadget, thus the new cycles created by the completions are even. Again by Claim 4.1, Claim 5.1 and Claim 6.1 we deduce that:

**Claim 6.4.** *The graph $\mathsf{G}^{s*}$ is a cubic planar bipartite graph.*

Moreover, Claim 6.2 and Claim 6.3 imply that:

**Claim 6.5.** *The graph $\mathsf{G}^s$ has a minimum $s$-club set of size $h$ if and only if the graph $\mathsf{G}^{s**}$ has a minimum $s$-club of size $h + 4k_2(s) + (12s - 6)k_3(s)$ when $s$ is odd, and of size $h + 4k_2(s) + (8s + 6)k_3(s)$ when $s$ is even.*

Then we have a reduction showing that $s$-Club-VD is NP-complete for cubic planar bipartite graphs.

## 7   $k$-PVC

Let $k \geq 3$. The hardness of $k$-PVC has been extensively studied, including on planar and/or bipartite graphs with bounded degree. It is known that 3-PVC is NP-complete for planar bipartite graphs of degree at most 4 [16], for cubic planar graphs of girth 3 [25], for subcubic bipartite graphs [3] and is APX-complete in bipartite graphs [16, 17]. Concerning $k$-PVC, the problem is NP-complete for each $k \geq 4$ [6] (and even for $k = 2$, which is Vertex Cover). Moreover, 4-PVC is NP-complete in cubic planar graphs [11] and APX-complete in cubic bipartite graphs [11].

In the case of graphs with girth at least $k + 1$, the problems $(k - 2)$-Club-VD and $k$-PVC are equivalent, since each $k$-path of the graph is a $P_k$.

In particular, 1-Club-VD (*i.e.* Cluster-VD) is equivalent to 3-PVC in bipartite graphs, since the girth of these graphs is at least 4. We then deduce from Theorem 3 with $s = 1$ the following new result, which improves the existing ones:

**Theorem 4.** 3-PVC *is NP-complete for cubic planar bipartite graphs.*

Moreover, recall that Theorem 1 and Corollary 4.5 are valid for graphs with girth larger than a given girth $g$. When we fix $g \geq k + 1$, we obtain a class of graphs for which $k$-PVC, with $k \geq 4$, is equivalent to $(k - 2)$-Club-VD. We therefore deduce the following statement:

**Theorem 5.** *Let* $k \geq 4$ *and* $g \geq k + 1$. *Problem* $k$-PVC *is NP-complete for subcubic planar bipartite graphs (when* $k$ *is odd), respectively for subcubic planar graphs (when* $k$ *is even) of girth larger than* $g$.

# 8 Conclusion

In this paper, we have shown that Cluster-VD and more generally $s$-Club-VD are NP-complete for cubic planar bipartite graphs. These problems, and Cluster-VD in particular, remain open for some well-known classes of graphs, among which for instance chordal graphs.

We have also deduced the NP-completeness of 3-PVC for cubic planar bipartite graphs. However, for $k \geq 4$, our results on $k$-PVC concern only subcubic (and not cubic) planar bipartite graphs. This is due to our gadgets $\mathsf{A}_s$ and $\mathsf{B}_s$ whose girth is four and which yield, therefore, instances $\mathsf{G}^{s*}$ on which $(k - 2)$-Club-VD and $k$-PVC are not equivalent. For $k \geq 4$, $k$-PVC for cubic planar bipartite graphs is therefore open.